\documentclass[longauth]{aa}

\usepackage{graphicx}
\usepackage{txfonts}

\usepackage{amsmath}
\usepackage{multirow}
\usepackage[section]{placeins}
\usepackage{hyperref}
\usepackage{mathtools}
\usepackage{xcolor}
\usepackage{soul}
\usepackage{ulem}
\usepackage{stfloats}

\makeatletter
\renewcommand*\aa@pageof{, page \thepage{} of \pageref*{LastPage}}
\makeatother

\hypersetup{colorlinks=true,allcolors=[rgb]{0,0,0.8}}

\newcommand{\simba}{{\sc Simba}}

\newcommand{\gizmo}{{\sc Gizmo}}
\newcommand{\gadget}{{\sc GADGET-3}}
\newcommand{\caesar}{{\sc Caesar}}

\newcommand{\grackle}{{\sc Grackle-3.1}}
\newcommand{\cosmosx}{{\sc COSMOS-XID+}}
\newcommand{\cosmos}{{\sc COSMOS}}
\newcommand{\cosmostwo}{{\sc COSMOS2020}}

\newcommand{\cosmosweb}{{COSMOS-Web}}
\newcommand{\disperse}{{\sc DisPerSE}}
\newcommand{\lephare}{{\sc Le Phare}}

\newcommand{\dfil}{\langle \mathrm{d}_{\mathit{fil}} \rangle}
\defcitealias{jego2026}{J26}

\usepackage{orcidlink}
\newcommand{\orcid}[1]{\orcidlink{#1}}

\begin{document}

   \title{The impact of cosmic filaments on starburst galaxies across cosmic times}
   \authorrunning{Jego, B., et al.}
   
   \author{Baptiste Jego\orcid{0009-0006-6399-7858}\inst{1},
          Matthieu Béthermin\orcid{0000-0002-3915-2015}\inst{1},
          Katarina Kraljic\orcid{0000-0001-6180-0245}\inst{1},
          Clotilde Laigle\orcid{0009-0008-5926-818X}\inst{2},
          Lingyu Wang\orcid{0000-0002-6736-9158}\inst{3, 4},
          Antonio La Marca\orcid{0000-0002-7217-5120}\inst{5, 6},
          Olivier Ilbert\orcid{0000-0002-7303-4397}\inst{7},
          Hollis B. Akins\orcid{0000-0003-3596-8794}\inst{8},
          Caitlin M. Casey\orcid{0000-0002-0930-6466}\inst{9, 10},
          Gavin Leroy\orcid{0009-0004-2523-4425}\inst{11},
          Ali Hadi\orcid{0009-0003-3097-6733}\inst{12},
          Jeyhan S. Kartaltepe\orcid{0000-0001-9187-3605}\inst{13},
          Anton M. Koekemoer\orcid{0000-0002-6610-2048}\inst{14},
          Henry Joy McCracken\orcid{0000-0002-9489-7765}\inst{2},
          Louise Paquereau\orcid{0000-0003-2397-0360}\inst{15},
          Jason Rhodes\orcid{0000-0002-4485-8549}\inst{16},
          Brant E. Robertson\orcid{0000-0002-4271-0364}\inst{17},
          Marko Shuntov\orcid{0000-0002-7087-0701}\inst{10, 18, 19},
          Greta Toni\orcid{0009-0005-3133-1157}\inst{20, 21, 22},
          Can Xu\orcid{0000-0002-8437-6659}\inst{23, 24, 25}
          }

   \institute{Observatoire Astronomique de Strasbourg, UMR 7550, CNRS, Université de Strasbourg, F-67000 Strasbourg, France\\ \email{baptiste.jego@astro.unistra.fr} \and
             Institut d’Astrophysique de Paris, UMR 7095, CNRS, Sorbonne Université, 98 bis boulevard Arago, 75014 Paris, France \and
             SRON Netherlands Institute for Space Research, Landleven 12, 9747 AD Groningen, The Netherlands \and
             Kapteyn Astronomical Institute, University of Groningen, Postbus 800, 9700 AV Groningen, The Netherlands \and
             European Space Agency/ESTEC, Keplerlaan 1, 2201 AZ Noordwijk, The Netherlands \and
             Leiden Observatory, Leiden University, Einsteinweg 55, 2333 CC Leiden, The Netherlands \and
             Aix Marseille Univ, CNRS, CNES, LAM, Marseille, France \and
             The University of Texas at Austin, 2515 Speedway Blvd Stop C1400, Austin, TX 78712, USA\and
             Department of Physics, University of California, Santa Barbara, Santa Barbara, CA 93106, USA \and
             Cosmic Dawn Center (DAWN), Denmark \and
             Institute for Computational Cosmology, Department of Physics, Durham University, South Road, Durham DH1 3LE, United Kingdom \and
             Department of Physics and Astronomy, University of California, Riverside, 900 University Avenue, Riverside, CA 92521, USA \and
             Laboratory for Multiwavelength Astrophysics, School of Physics and Astronomy, Rochester Institute of Technology, 84 Lomb Memorial Drive, Rochester, NY 14623, USA \and
             Space Telescope Science Institute, 3700 San Martin Dr., Baltimore, MD 21218, USA \and
             Department of Space, Earth and Environment, Chalmers University of Technology, SE-412 96 Gothenburg, Sweden\and
             Jet Propulsion Laboratory, California Institute of Technology, 4800 Oak Grove Drive, Pasadena, CA 91001, USA \and
             Department of Astronomy and Astrophysics, University of California, Santa Cruz, 1156 High Street, Santa Cruz, CA 95064, USA \and
             Niels Bohr Institute, University of Copenhagen, Jagtvej 128, DK-2200, Copenhagen, Denmark\and
             University of Geneva, 24 rue du Général-Dufour, 1211 Genève 4, Switzerland\and
             University of Bologna, Department of Physics and Astronomy “Augusto Righi” (DIFA), Via Gobetti 93/2, I-40129 Bologna, Italy \and
             INAF - Osservatorio di Astrofisica e Scienza dello Spazio, Via Gobetti 93/3, I-40129 Bologna, Italy \and
             Zentrum f\"ur Astronomie, Universit\"at Heidelberg, Philosophenweg 12, D-69120 Heidelberg, Germany \and
             School of Astronomy and Space Science, Nanjing University, Nanjing, Jiangsu 210093, China \and
             Key Laboratory of Modern Astronomy and Astrophysics, Nanjing University, Ministry of Education, Nanjing 210093, China \and
             Kavli Institute for the Physics and Mathematics of the Universe (WPI), The University of Tokyo, Kashiwa, Chiba 277-8583, Japan
            }

    \date{Received 25 February 2026 ; Accepted 29 May 2026}

   \abstract
   {
    Cosmological simulations suggest that various galaxy properties depend on their location within the cosmic web. Yet direct observational evidence of the dependence of star formation activity on distance to filaments remains scarce, and is missing at $z\gtrsim1$. We investigate how starburst, main-sequence (MS), and quenched galaxies are distributed with respect to cosmic web filaments, and how this distribution evolves with redshift.
    We first use the \simba~cosmological hydrodynamical simulation to predict the redshift evolution of the mean distance to the closest filament from $z=3$ to $z=0$ for different galaxy populations, after removing stellar-mass dependencies. We then measure the corresponding signal in the \cosmos~field, using \cosmostwo~and \cosmosweb~data, where accurate photometric redshifts enable a reconstruction of the projected cosmic web from $z=2$ to $z=0.5$, and starbursts are identified through far-infrared spectral energy distribution fitting. 
    In agreement with the results from \simba, starburst galaxies are found closer to filaments at $z>1$ and at larger distances at $z<1$, MS galaxies occupy intermediate environments with little evolution, and quenched galaxies show progressively shorter distances to filaments toward low redshift, with a crossing between starburst and MS populations around $z \sim 1$. In \cosmosweb, the relative evolution in the average distance to filaments between starburst and MS galaxies is detected at a significance level of at least $5\sigma$.
    We show that a minimal toy model in which the only environmental ingredient is the sSFR-filament distance modulation measured in simulations is sufficient to reproduce the observed differential evolution of the average filament distance between starburst and MS galaxies.
    These results show evidence for a link between the large-scale environment and the star formation activity of galaxies, as predicted by simulations, from $z=2$ down to $z=0.5$.
   }


   \maketitle

\section{Introduction}\label{sec:intro}

Understanding galaxy formation and evolution remains one of the key challenges in astrophysics, particularly when it comes to disentangling the role played by the large-scale environment. While the mass assembly of galaxies broadly traces the growth of dark matter haloes \citep{1984Natur.311..517B, 1996ApJ...462..563N}, it is also strongly influenced by internal baryonic processes and interactions with the cosmic web. This web-like structure, composed of voids, walls, filaments, and nodes, naturally emerges from the gravitational collapse of initial density perturbations, as described by the Zel’dovich approximation \citep{1970A&A.....5...84Z}, and forms the frame within which galaxies grow and evolve \citep{1996Natur.380..603B}. 

Filaments and nodes, in particular, are expected to be linked to galaxy evolution through their enhanced densities and the anisotropic tidal fields, which regulate halo growth and the directions of matter inflows \citep[e.g.][]{Pichon2011, Codis2012, Dubois2014, kraljic2019,  Kraljic2020}. In this framework, filaments act as preferential channels for the accretion of gas and dark matter onto haloes, funnelling material from under-dense regions towards high-density nodes and modulating galaxy mass assembly histories \citep[e.g.][]{Keres2005, Dekel2009, Welker2014, Rost2021}. 

Thus, filaments are expected to modulate galaxy growth and star formation through a range of mechanisms at different scales.
On cosmological scales, where filaments are traced from the galaxy distribution, observational studies consistently find that galaxies closer to filaments tend to be more massive, redder, and less star-forming even after controlling for stellar-mass dependencies, and that their satellite fraction increases with proximity to filaments \citep[e.g.][but see also \citealt{okane2024}]{Martinez2016, malavasi2017, kraljic2018, laigle2018, Winkel2021, mondelin2025}. These effects are also predicted by modern large-scale hydrodynamical simulations \citep[e.g.][]{Hasan_2023, 10.1093/mnras/stae667, Zakharova2026}.

At cluster scales, observational analyses suggest that galaxies infalling along filaments exhibit suppressed star-formation and distorted morphologies consistent with pre-processing before entering the cluster core \citep[e.g.][]{Sarron2019}. Simulations indicate that this behaviour depends on the physical nature of the filamentary structures considered: large-scale filaments traced by galaxies are environmentally heterogeneous, hosting individual infalling galaxies, back-splash systems, and groups, and promote multiple pre-processing pathways prior to accretion \citep[e.g.][]{Kuchner2022, Zakharova2025}, whereas intra-cluster filaments traced by the gas density field may instead mitigate strangulation and ram-pressure stripping, leading to weaker star-formation suppression \citep{Kotecha2022}.

Overall, these patterns are indicative of enhanced pre-processing, tidal interactions, and potentially suppressed gas accretion leading to quenching of galaxies in the densest environments in both observations \citep[e.g.][]{10.1093/mnras/stw3127, malavasi2017, kuutma2017, kraljic2018, laigle2018} and simulations \citep[e.g.][]{Hasan_2023, 10.1093/mnras/stae667}.
In particular, simulations indicate that quenching does not always stem from suppressed gas accretion alone, as at filament edges, where vorticity and angular-momentum inflows peak, star formation can be hindered even though gas supply is not strictly reduced \citep{song2020}. These environmental effects are intimately tied to assembly bias, wherein galaxy properties at fixed mass correlate with the larger-scale environment \citep[e.g.][]{10.1111/j.1365-2966.2004.07733.x, 2005MNRAS.363L..66G, wechsler2006, Borzyszkowski2017, Paranjape2018, musso2018}.

Whether the large-scale environment modulates star formation within star-forming galaxies (SFG) is still an open question. This question has only recently started to be addressed. In their recent study, \citet[][\citetalias{jego2026} hereafter]{jego2026} studied deviation in the star formation rate (SFR), specific SFR (sSFR, defined as the SFR divided by stellar mass), gas fractions and depletion timescales of SFG in the \simba~simulation \citep{Dave2019}, after removing stellar-mass dependencies, as a function of the distance to cosmic filaments for redshifts between $z=0$ and $z=3$. They showed redshift-dependent modulations of these properties as a function of distance to filaments. At low redshift, these trends appear to be primarily linked to satellite-related processes, while at high redshift they are best explained by accretion-driven effects, with no evidence for a dominant contribution from mergers. However, current observational datasets are not suited to reproduce such a detailed analysis.

Dust-rich starburst galaxies offer a particularly powerful probe of how the large-scale environment may modulate star formation. These systems experience short-lived episodes of star formation well above the main sequence and typically host large dust reservoirs, such that their star formation rates are most reliably constrained through mid- to far-infrared (IR) emission and full spectral energy distribution (SED) modelling. While FIR-based SFR indicators are not strictly required for moderately star-forming galaxies, they become essential for capturing the heavily obscured star formation that dominates the emission of starbursts. Their high FIR luminosities ($L_{\rm IR}$), often associated with mergers or rapid gas inflows, therefore make starburst galaxies especially sensitive tracers of environmental triggering or suppression mechanisms across cosmic time \citep[e.g.][]{Elbaz2011, Rodighiero2011, Sargent2012, Schreiber2015, Faisst2025}.

Robustly identifying SFG and measuring their SFRs, and especially those of dust-obscured starbursts, requires far-IR (FIR) observations as they trace dust-reprocessed emission from young stars and provide extinction-free measures of star formation activity. FIR surveys from observatories such as \textit{Herschel} \citep{Pilbratt2010} and \textit{Spitzer} \citep{Werner2004} have revealed the existence of an extreme population of starbursts with elevated SFRs, constituting only $\sim10\%$ of the total SFR density \citep{Rodighiero2011}, often undetectable or misclassified in ultraviolet (UV) or optical surveys due to heavy dust obscuration \citep[e.g.][]{Heinis2013b, fudamoto2020}. Additionally, more recent works have demonstrated that standard SED fitting techniques, when relying on UV-to-optical data alone, often fail to recover the full extent of cold dust and obscured star formation in dust-rich galaxies \citep[see e.g.][]{Noll2009, Casey2014, Leja2017, li2024}, highlighting the need for FIR and sub-mm data from \textit{Herschel} and ALMA \citep{alma2015} to accurately constrain SFRs and dust properties \citep[e.g.][]{Buat2019}. 

On the other hand, identifications of cosmic-web filaments from galaxy positions in observational datasets are becoming more and more reliable. Over the past two decades, the three-dimensional structure of the cosmic web has progressively been unveiled by large spectroscopic surveys such as SDSS \citep{SDSS2000}, which revealed the filamentary network of galaxies in the local Universe \citep{deLapparent1986, Tegmark2004}. At higher redshifts, dedicated photometric and spectroscopic programs such as VIPERS \citep{Guzzo2013, malavasi2017} and \cosmos~\citep{cosmos2007, laigle2018} have extended this picture out to $z\sim 1$.

In this work, we build upon the study by \citetalias{jego2026} that focused primarily on SFG within the \simba~simulation. To go beyond simulations and study the link between proximity to cosmic filaments traced by galaxy distributions and the star formation activity, we adopt a complementary and simpler approach to measure the mean distance to filaments for quenched passive galaxies, star-forming main-sequence (MS) galaxies, and starburst galaxies as a function of redshift. 
We first establish predictions using the \simba~cosmological hydrodynamical simulation. We quantify how the mean distance to filaments $\dfil$ evolves with redshift from $z=0$ to $z=3$ for starburst, MS, and quenched galaxies. We then confront these predictions with real data from the \cosmos~field. Using the \cosmosx~FIR catalogues \citep{cosmosxid2024}, which offer reliable deblended flux measurements from \textit{Herschel} imaging, we identify robust samples of starburst galaxies based on their $L_{\rm IR}$. Combined with photometric redshifts from \cosmostwo~\citep{cosmos2020} and \cosmosweb~\citep{cosmosweb2023, cosmosweb2025}, these datasets enable a measurement of the mean projected distance to filaments for each galaxy type as a function of redshift. Since the \cosmos~field relies on photometric redshifts, any cosmic-web reconstruction is by construction two-dimensional \citep[e.g.][]{laigle2018}. We thus adopt a 2D framework for both the observations and the \simba~simulation, ensuring a consistent comparison, with the 3D reconstruction in \simba~used only for interpretation.

The paper is structured as follows. In Sect.~\ref{sec:simba} we describe the simulated data extracted from the \simba~simulation and the resulting predictions. In Sect.~\ref{sec:sample}, we detail our method to build a robust selection of starburst galaxies in the \cosmos~field and test our predictions with observational data. Our results are then presented in Sect.~\ref{sec:obs}. We discuss these results in Sect.~\ref{sec:intp} and our conclusions are summarised in Sect.~\ref{sec:ccl}. 
Additional material is provided in the Appendices. Appendix~\ref{appA} gives details on the number of galaxies at different snapshots in \simba. Appendix~\ref{appB} presents examples of FIR SED fitting. Appendix~\ref{appC} quantifies the impact of photometric masks on cosmic web reconstructions. Appendix~\ref{appD} shows measurements based on the distance of galaxies to cosmic web nodes. Appendix~\ref{appE} presents validation tests with randomised FIR luminosities. Finally, Appendix~\ref{appF} provides additional information on the modulation of sSFR by the distance to cosmic filaments in \simba.

We adopt a flat $\Lambda$CDM cosmology consistent with \textit{Planck} constraints \citep{Planck2016}, as implemented in \simba: $\Omega_{m}=0.3$, $\Omega_{\Lambda}=0.7$, $\Omega_{b}=0.048$, $H_{0}=68\,\rm km\,s^{-1}\,Mpc^{-1}$ (the \cosmostwo~and \cosmosweb~catalogues use $H_{0}=70\,\rm km\,s^{-1}\,Mpc^{-1}$), $\sigma_{8}=0.82$, and $n_{s}=0.97$. We refer to the logarithm in base ten as log.

\section{Predictions from the \simba~simulation}\label{sec:simba}

We first use the \simba~cosmological hydrodynamical simulation to reconstruct 2D cosmic webs from projected galaxy positions (Sect.~\ref{ssec:simba_cw}). Then, we predict the distribution of starburst, MS, and quenched galaxies relative to cosmic filaments (Sect.~\ref{ssec:simba_evol}). These predictions allow us to isolate environmental effects on star formation and provide a direct comparison to observational measurements of filament distances.

\subsection{Synthetic data}\label{ssec:simba_cw}

The \simba\footnote{\href{http://simba.roe.ac.uk/}{http://simba.roe.ac.uk/}} simulation suite is a set of cosmological hydrodynamical simulations designed to study galaxy formation and evolution across cosmic time \citep[see][ for a full description of the simulation]{Dave2019}. It incorporates advanced models for black hole growth, active galactic nuclei (AGN) feedback, and dust physics, and uses the Meshless Finite Mass hydrodynamics method from the \gizmo~code \citep{Hopkins2015}, combined with gravity calculated via the \gadget~tree-particle-mesh solver \citep{Springel2005}. The \simba~simulation used in this work evolves $1024^{3}$ dark matter and gas particles in a $(100~h^{-1}\rm\,cMpc)^3$ volume, reaching a mass resolution of $\sim10^{7}~\rm M_{\odot}$ per particle. It tracks galaxy formation from $z=99$ to $z=0$ in a self-consistent cosmological context. Radiative cooling is modelled with \grackle~\citep{Smith2017}, while star formation follows a Schmidt law based on the molecular hydrogen density, and stellar feedback is implemented via kinetic winds with mass loading factors calibrated on FIRE simulations \citep{Muratov2015,Angles-Alcazar2017}. Black hole growth is split between torque-limited and Bondi accretion modes \citep{Angles-Alcazar2015}, with AGN feedback occurring in radiative and jet modes depending on the Eddington ratio. \simba~also includes a detailed, on-the-fly model of dust production, growth, and destruction. Halos and galaxies are identified using the \caesar~package \citep{2011ApJS..192....9T}. 

We use the main \simba~runs, focusing on full physics snapshots at $z=0$, 0.5, 1, 1.25, 1.5, 2, 2.5, and 3 and using the catalogue extracted with \caesar~to infer galaxy properties. We use the publicly available software \disperse\footnote{\href{https://www2.iap.fr/users/sousbie/web/html/indexd41d.html}{https://www2.iap.fr/users/sousbie/web/html/indexd41d.html}} \citep{Sousbie2011, SousbiePK2011} to extract the skeleton of the cosmic web from the spatial distribution of all galaxies in each snapshot. \citetalias{jego2026} employed \disperse~to reconstruct the cosmic web in three dimensions from the full galaxy distribution, adopting a persistence threshold of $3\sigma$ and a single smoothing step. In the present work, we adopt a persistence threshold of $2.5\sigma$ for cosmic web reconstructions in both three and two dimensions. This adjustment does not significantly affect our analysis but facilitates a more straightforward application of the method to observational datasets. Moreover, we perform the reconstruction in two dimensions after projecting the three-dimensional galaxy positions over $80\,\rm cMpc$-deep slices. The projection is performed along one spatial axis. We have verified that repeating the analysis along the other simulation axes yields statistically consistent results, with no impact on the inferred trends. This setup allows a direct comparison with observational analyses performed in projected space.

The galaxy number density in the simulation evolves from $\sim 5\,\rm gal/cMpc^{2}$ at $z=0$ to $\sim 0.5\,\rm gal/cMpc^{2}$ at $z=3$ in projection, which sets the typical scale of the reconstructed filaments. In this context, the adopted persistence threshold primarily controls the significance of the detected structures rather than their characteristic scale. We verified that varying the persistence threshold within the range commonly used in the literature ($2$-$3\sigma$) does not affect our predictions, although it changes the number and typical size of filaments. We note that the reconstruction of the cosmic web depends on the sampling of the underlying galaxy distribution. Differences in galaxy number density between simulations and observations may therefore affect the detailed geometry of the extracted skeleton. 
However, we stress that our goal is not to establish a one-to-one correspondence between individual filamentary structures identified in \simba~and in \cosmos, but rather to compare relative statistical trends of galaxy populations with respect to their local cosmic web environment. To explicitly test this effect, we have performed a downsampling of the \simba~galaxy catalogue to match the number density of the \cosmos~sample at each redshift. Repeating the full analysis on this degraded sample does not qualitatively change the measured average trends of the relative ordering between galaxy populations.
In addition, we note that the \cosmos~field probes a limited cosmological volume and is therefore affected by cosmic variance and survey geometry effects. These cannot be reproduced by a simple downsampling of the simulation, and would require the construction of realistic lightcones with matched selection functions. Such an approach is beyond the scope of this work. Our analysis therefore focuses on relative, statistical trends within each dataset rather than on a direct comparison of individual filamentary structures.

\subsection{Evolution of the distance to cosmic filaments: prediction}\label{ssec:simba_evol}

\begin{figure}[ht]
        \centering
        \includegraphics[width=\columnwidth]{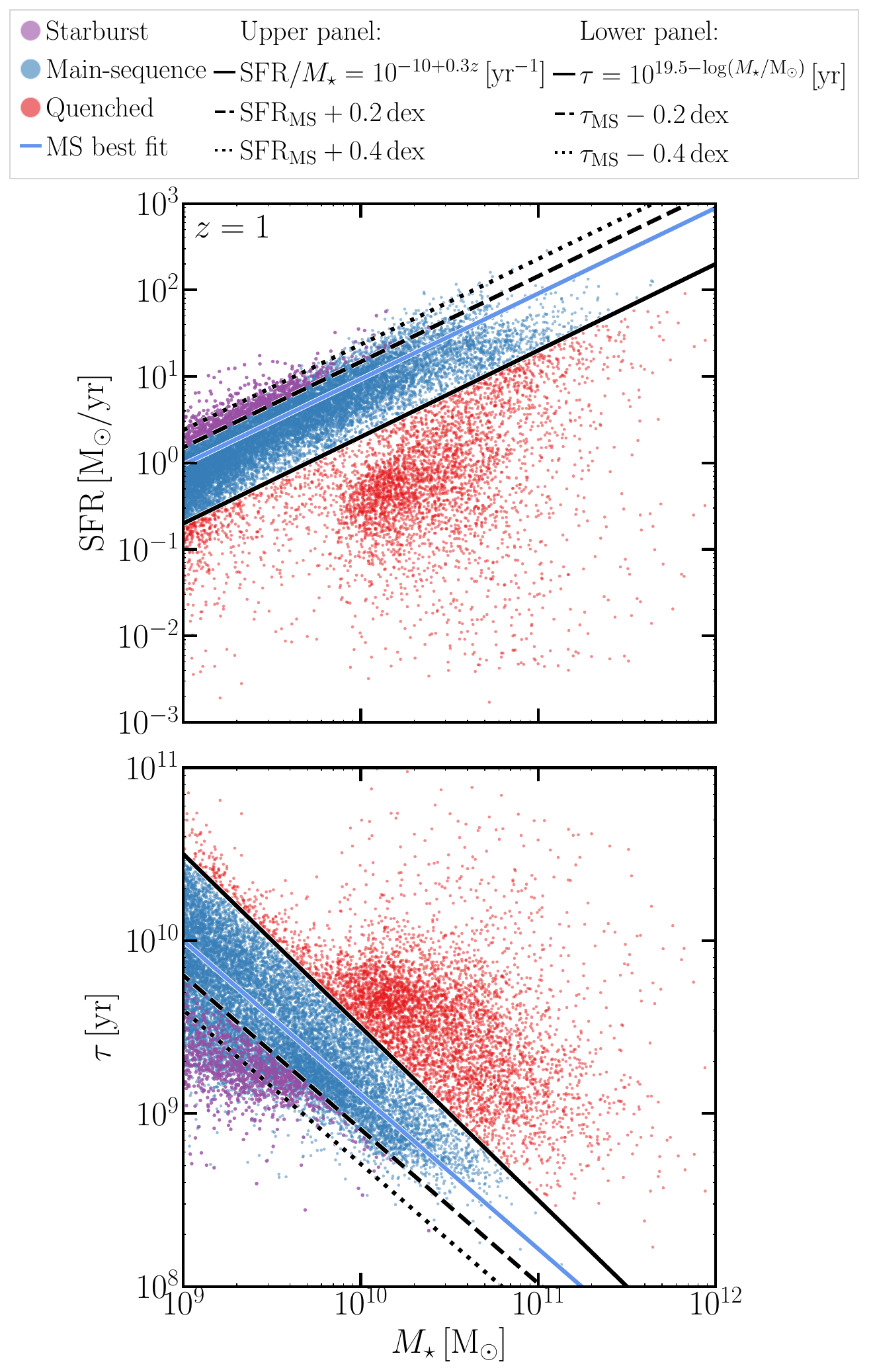}
        \caption{Upper panel: Separation of starbursts (purple), MS galaxies (blue) and quenched galaxies (red) in \simba~using their locus in the $\mathrm{SFR} - M_{\star}$ diagram at $z=1$ with $\log(\rm SFR_{SB}/SFR_{MS})>0.2\, dex$ and $\log(\rm \tau_{SB}/\tau_{MS})<-0.2\, dex$. Lower panel: Separation in the $\tau - M_{\star}$ diagram with the same starburst-selection criteria. The blue solid lines indicate the best-fit MS relations, i.e. the SFR-$M_{\star}$ and $\tau$-$M_{\star}$ relations for SFG in the upper and lower panels, respectively.}
        \label{fig:selection}
\end{figure}

\begin{figure*}
        \centering
        \includegraphics[width=\textwidth]{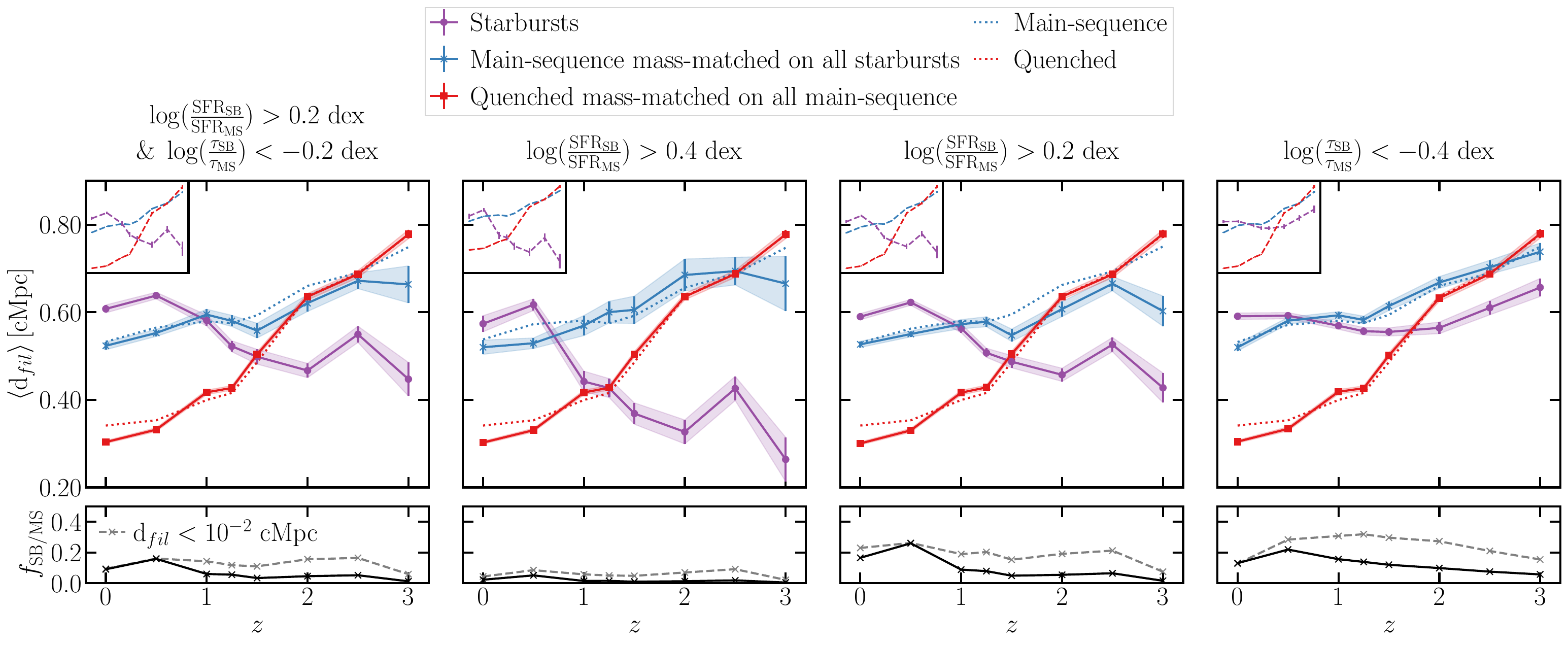}
        \caption{Upper panels: Mean distance to the cosmic web filaments, measured in projection (2D), for starburst galaxies (purple), MS galaxies mass-matched to starbursts (blue), and quenched galaxies mass-matched to the complete main sequence sample (red) as a function of redshift for four different selection criteria in \simba, with our reference case on the left. The dotted lines show the same quantities without mass-matching. The upper left boxes show the same trends for the median values. Lower panels: Fraction of starburst galaxies with respect to MS galaxies as a function of redshift. Starbursts are, on average, closer to filaments than MS galaxies at high redshift but exceed the MS distance at $z\simeq1$. The fraction of SFG rises with redshift and stabilises around $z\simeq1.5$, while the starburst fraction remains roughly constant at $\sim10\%$, with a modest enhancement for galaxies near filaments at $z>1$.}
        \label{fig:d_z_prediction}
\end{figure*}

\begin{figure}[ht]
        \centering
        \includegraphics[width=\columnwidth]{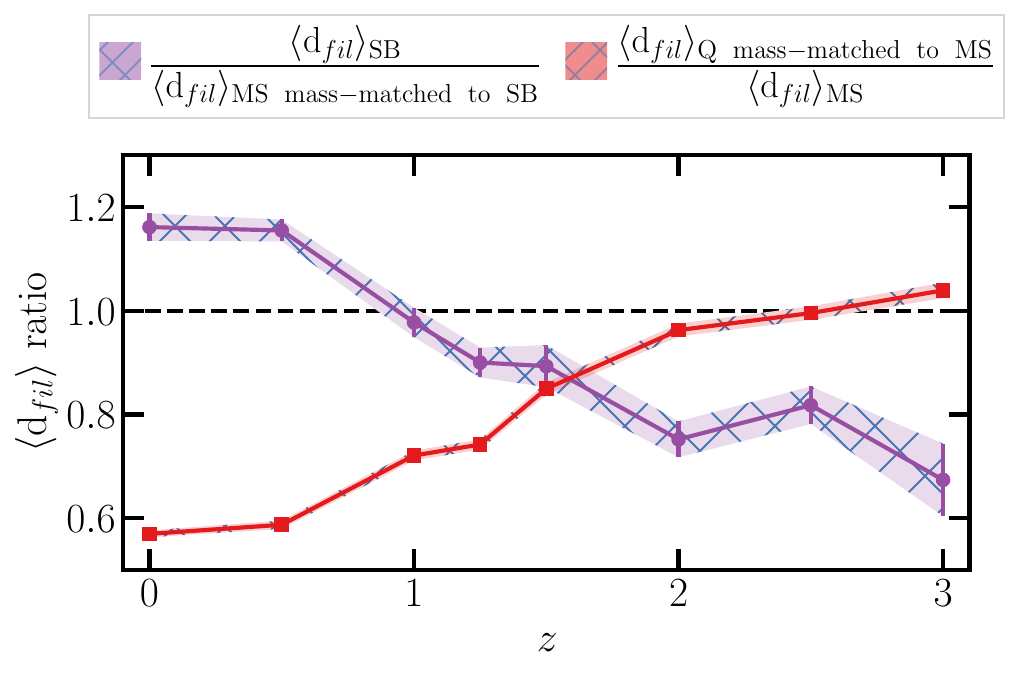}
        \caption{Ratio of the mean projected (2D) distance to cosmic filaments for two galaxy populations as a function of redshift in \simba~with $\log(\rm SFR_{SB}/SFR_{MS})>0.2\, dex$ and $\log(\rm \tau_{SB}/\tau_{MS})<-0.2\, dex$. The purple curve shows the ratio between starbursts and MS galaxies mass-matched to starbursts (solid purple line and solid blue line in Fig.~\ref{fig:d_z_prediction}). The red curve shows the ratio between quenched galaxies mass-matched to the MS sample and MS galaxies (solid red line and dotted blue line in Fig.~\ref{fig:d_z_prediction}). At late cosmic times, starbursts are, on average, located farther from filaments than their mass-matched MS counterparts. On the other hand, quenched galaxies lie, on average, closer to filaments than mass-matched MS galaxies, and the contrast between the two increases towards lower redshift.}
        \label{fig:d_z_prediction_ratios}
\end{figure}

We aim to study the distribution of starburst and MS galaxies as a function of distance to filaments to isolate how star formation activity responds to large-scale environmental effects beyond the underlying stellar mass dependence. To do so, in \simba, we select galaxies following a similar procedure to \citetalias{jego2026}, i.e. with stellar masses above $10^{9}\,\rm M_{\odot}$, and we classify systems between star-forming and quenched based on their sSFR, with SFG having $\mathrm{sSFR} > 10^{-10 + 0.3\times \mathrm{min}(z, 2)}\,\rm yr^{-1}$, where $\mathrm{min}(z, 2)$ denotes the minimum of $z$ and 2. This selection is a slight adaptation of the prescription from \citet{Dave2019}, providing a robust definition of SFG and accounting for the evolution of the mean SFR from $z=0$ to $z=3$. Additionally, the star-forming main sequence is then constructed by fitting the SFR - $M_{\star}$ to the SFG population, ensuring that it is not biased by transitional or passive galaxies. For each galaxy, we compute the offset of its SFR from the main sequence at its stellar mass and redshift. 

To identify starburst galaxies, we define a fiducial selection and explore several alternative criteria to assess the robustness of our results. Figure~\ref{fig:selection} illustrates these selection criteria in the $\mathrm{SFR}-M_{\star}$ and $\tau-M_{\star}$ planes at $z=1$, where $\tau=M_{\rm gas}/\mathrm{SFR}$ is the gas depletion time, highlighting the separation between starburst, MS, and quenched galaxies for different thresholds adopted in this work.
Our reference definition requires that a galaxy satisfies both $\log(\mathrm{SFR}_{\rm SB}/\mathrm{SFR}_{\rm MS}) > 0.2\,\rm dex$, and $\log(\tau_{\rm SB}/\tau_{\rm MS}) < -0.2\,\rm dex$.
This dual requirement ensures that selected starbursts are both forming stars at an elevated rate and consuming their gas reservoir more rapidly than typical SFG.

In addition to this fiducial definition, we consider three alternative selections with a pure SFR-based criterion with thresholds of $0.2\,\rm dex$ and $0.4\,\rm dex$ above the main sequence, and a depletion-time-only criterion with $\tau$ at least $0.4\,\rm dex$ below the main sequence.
The main sequence and depletion time relations are always derived from the star-forming population, ensuring that the reference values for both SFR and $\tau$ reflect the properties of actively star-forming systems at each mass and redshift. This approach follows observational strategies and maintains consistency with empirical studies. The precise value of the depletion time offset used to define starbursts does not significantly affect our results, as long as it remains within the $0.4 \pm 0.2\,\rm dex$ range reported by \citet{DG2020}. We choose a relatively low value, as the scatter around the median relation is narrow in \simba, and because depletion times are only available in the simulation, whereas the observational analysis will rely on a more stringent SFR-based selection.
These alternative definitions are not used for the main analysis, but to verify that our results do not depend on the choice of the criterion used to select starbursts. The corresponding galaxy numbers for each redshift snapshot and selection are provided in Appendix~\ref{appA}, Table~\ref{tab:ngals}.

To study the evolution of the average distance to filaments, $\dfil$, for our different galaxy selections while isolating environmental effects beyond stellar mass, we construct mass-matched samples of MS galaxies for each starburst population, and of quenched galaxies for each MS population at each redshift. These mass-matched samples ensure that observed differences in $\dfil$ reflect environmental trends rather than mass-driven effects. For each starburst galaxy, we iteratively select at the same redshift the MS galaxy with the closest stellar mass, removing each matched MS galaxy from the pool to enforce a one-to-one correspondence. In cases where several starburst galaxies share the same nearest neighbour, the match is assigned to the pair with the smallest mass difference and the second starburst is assigned to the next nearest MS galaxy. We have verified that the mass distributions between the mass-matched samples are the same for each \simba~snapshot. For reference, there are twice as many galaxies selected as starbursts with $\log(\tau_{\rm SB}/\tau_{\rm MS}) < -0.4\,\rm dex$ as the only criterion than with both $\log(\rm SFR_{\rm SB}/SFR_{\rm MS}) > 0.2\,\rm dex$ and $\log(\tau_{\rm SB}/\tau_{\rm MS}) < -0.2\,\rm dex$. We have also verified that the mass matching is extremely precise, with typical differences between matched pairs of $\Delta \log (M_{\star}/\rm M_{\odot}) \sim 10^{-4}$-$10^{-3}$ dex.

Importantly, the mass-matching procedure is designed to enable two distinct comparisons, both using the MS population as a reference. Specifically, starburst galaxies are compared to mass-matched MS galaxies, while MS galaxies are independently compared to mass-matched quenched galaxies. At no point are the starburst and quenched populations directly compared. We note that we choose to match on the stellar mass to ease the comparison with observations and their interpretation (see Sect.~\ref{sec:obs}). Alternatively, the matching procedure can be performed on the halo mass. We verified that such an approach does not alter our conclusions. 
Similarly, excluding galaxies in massive clusters, i.e. with host halo masses $\geqslant 10^{13}\,\rm M_{\odot}$ leads to only minor quantitative changes (at the $\sim1-2\sigma$ level) and does not change our predictions and conclusions. This is expected, as galaxies residing in massive halos represent only a small fraction of our star-forming sample at all redshifts (a few-percent), and their relative contribution decreases with increasing redshift.

Our predictions for the evolution of $\dfil$ with redshift, from $z=3$ to $z=0$, for the four different selection criteria are presented in the upper panels of Fig.~\ref{fig:d_z_prediction}, where all distances to filaments are measured from cosmic web reconstructions performed from projected 2D galaxy positions, in order to mimic observational conditions. Although different starburst selection criteria yield different absolute variations of $\dfil$ as a function of redshift for the three populations, they show similar relative trends between populations. At high redshift ($z\gtrsim2$), MS galaxies (in blue in Fig.~\ref{fig:d_z_prediction}) and their quenched counterparts (red) have comparable $\dfil$ values, and both populations exhibit decreasing $\dfil$ toward lower redshift, with the steepest decline for the quenched population. Starburst populations (purple) exhibit a distinct behaviour: apart from the selection based on the gas depletion timescale alone ($4^{\rm th}$ panel), they are preferentially located closer to cosmic filaments at high redshift. Toward lower redshift, their mean distance to filaments remains roughly constant or increases slightly for all selection criteria. 

The bottom panels of Fig.~\ref{fig:d_z_prediction} show the fraction of starbursts relative to the full MS sample as a function of redshift. The solid black curves correspond to the total fraction, while the dashed grey curves show the same quantities restricted to galaxies located very close to filaments ($\dfil < 10^{-2}\,\mathrm{cMpc}$). The starburst-MS fraction remains around $\sim10\%$ with only modest redshift dependence. Interestingly, for all selections, this fraction is higher among galaxies located close to filaments at $z>1$, suggesting that intense star formation episodes are relatively more common in dense filamentary environments at early times. This is in agreement with the excess in star formation rate at small distances to filaments at high redshift seen in \citetalias{jego2026}. We note that the fraction of SFG relative to the full sample of galaxies, which is the same for all selection criteria, is stable at high redshifts, down to $z\simeq1.5$ where it decreases with decreasing redshift. It decreases from $70-80\%$ to about $40\%$ at $z=0$. The same fraction from only galaxies close to filaments is similar at $z\gtrsim2$, but decreases steadily toward lower redshift, reaching $\sim20\%$ by $z=0$.

Figure~\ref{fig:d_z_prediction_ratios} presents the ratio of the mean distance to cosmic filaments between two galaxy populations (starbursts compared to MS and quenched compared to MS) as a function of redshift, using our default starburst selection criterion ($\log(\rm SFR_{SB}/SFR_{MS})>0.2\, dex$ and $\log(\rm \tau_{SB}/\tau_{MS})<-0.2\, dex$). The purple curve shows the ratio between starbursts and MS galaxies mass-matched to starbursts, while the red curve shows the ratio between quenched galaxies mass-matched to the MS population and the MS galaxies themselves. Ratios above unity indicate that the numerator population lies, on average, farther from filaments than the denominator. This representation highlights population-level contrasts more directly than the absolute distances shown in Fig.~\ref{fig:d_z_prediction}, isolating the relative environmental positioning of starbursts and quenched galaxies with respect to their mass-matched MS counterparts. The ratio between starburst and MS galaxies rises toward lower redshift and crosses unity around $z\simeq1$, indicating that at late cosmic times starbursts are, on average, located farther from filaments than MS galaxies of similar mass. In contrast, the ratio between quenched and MS galaxies declines from unity at $z=3$, showing that quenched galaxies lie progressively closer to filaments than mass-matched MS galaxies, with the difference increasing toward low redshift. This figure therefore provides a clear, quantitative prediction of how environmental segregation between populations evolves over cosmic time, which we can test with observational samples (see Sect.~\ref{ssec:obs_res}). This ratio-based statistic is indeed well-suited for comparison with observational measurements, as it is expected to be less sensitive to redshift-dependent observational effects than the mean distance measured for each population. In practice, with observational data, because both galaxy populations are selected within the same redshift slices, systematics such as redshift uncertainties, slice thickness, and incompleteness should affect them in a similar way and thus largely cancel out in the ratio.

\section{Building a starburst sample from observations}\label{sec:sample}

In this section, we construct a robust sample of starburst galaxies from the \cosmos~field. Using deep multi-wavelength photometric catalogues and FIR fluxes (Sect.~\ref{ssec:sample_cats}), we identify highly star-forming systems and define comparison samples of MS and quenched galaxies (Sect.~\ref{ssec:sample_sel}).

\subsection{Photometric catalogues}\label{ssec:sample_cats}

The \cosmos~field \citep{cosmos2007} is one of the most intensively studied extragalactic deep fields, combining deep, multi-wavelength imaging over $2\,\rm deg^{2}$ from a wide range of ground- and space-based observatories. The latest version of the catalogue combining this large set of available data is published in the \cosmostwo~photometric catalogue \citep{cosmos2020}. It assembles imaging across 38 bands from the UV to the mid-IR ($0.1 < \lambda < 10~\mu\rm m$), including data from HST, \textit{Subaru}/HSC, CFHT/CLAUDS, UltraVISTA, \textit{Spitzer}/IRAC, and GALEX.

\cosmostwo~provides two parallel measurements of photometry and derived galaxy properties. The {\sc Classic} catalogue is based on aperture photometry performed on PSF-homogenised images, providing consistent colours across all bands. The {\sc Farmer} catalogue \citep{Weaver2023} uses forward modelling with {\tt The  Tractor} \citep{tractor2016}, fitting galaxy profiles on the native-resolution images to improve deblending and photometric accuracy in crowded fields. Both methods yield consistent estimates of photometric redshifts, stellar masses, and rest-frame magnitudes, with physical properties derived using the \lephare~code \citep{1999MNRAS.310..540A,2006A&A...457..841I}. In this work, we adopt the {\sc Farmer} catalogue which provides improved flux measurements and colour fidelity compared to traditional aperture photometry, especially in crowded or blended regions such as the \cosmos~field \citep[see][for a comparison between the two methods in the \cosmos~field]{cosmos2020}. The combination of deep near-IR imaging from UltraVISTA DR6 and \textit{Subaru}/HSC-SSP PDR2 allows reliable photometry down to $i \sim 27$ and robust redshift estimation out to $z \sim 6$, with a typical photometric redshift precision of $\sigma_{\Delta z / (1+z)} \sim 0.02$ for SFG at $z < 3$.

In addition, the \cosmosweb~survey \citep{cosmosweb2023} is a JWST Cycle~1 Treasury program covering $\sim$0.54~deg$^2$ at the center of the \cosmos~field, in four NIRCam bands (F115W, F150W, F277W, F444W) and $\sim$0.2~deg$^2$ in MIRI/F770W, reaching very deep near- and mid-IR sensitivities. \cosmosweb~enables the detection and morphological characterisation of galaxies out to $z \sim 10$. Combined with the full \cosmos~photometric dataset, it provides improved photometric redshifts and more reliable stellar mass estimates at high redshift ($z\gtrsim1.5$), as well as detailed morphological information enabled by JWST \citep{cosmosweb2025}. In this work, the primary advantage of \cosmosweb~lies in the improved photometric redshift constraints at high redshift, which are crucial for cosmic-web studies.

From these catalogues, we extract for each galaxy its right ascension and declination, photometric redshift, redshift probability density function, and stellar mass, which define the basis of our galaxy sample. These quantities are used throughout our analysis for sample selection, environment measurements, and comparisons with theoretical predictions.

\subsection{Starburst far-infrared selection}\label{ssec:sample_sel}

\begin{figure}[ht]
        \centering
        \includegraphics[width=\columnwidth]{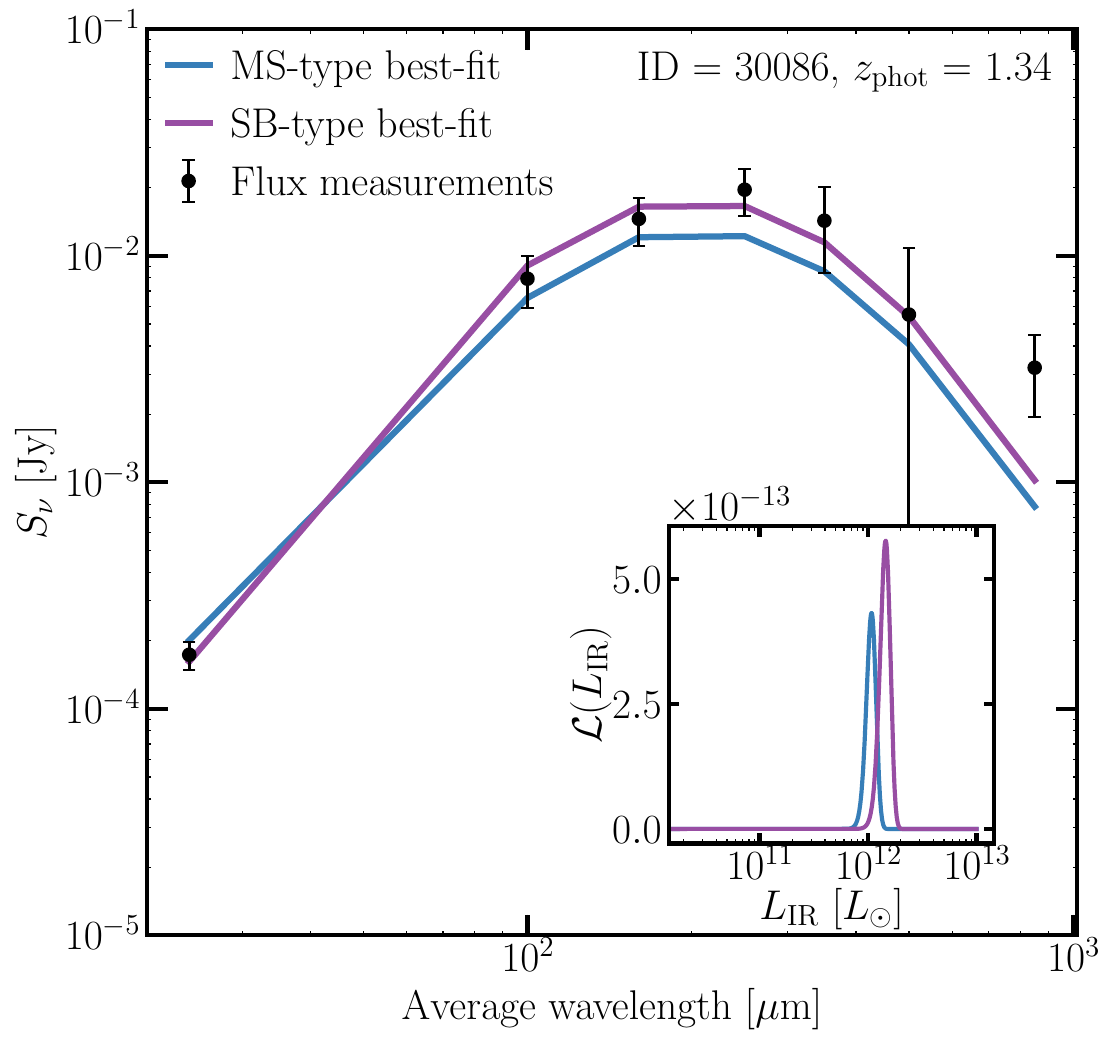}
        \caption{Example of FIR SED fitting for one galaxy in our sample. Black points show the observed photometric fluxes as a function of wavelength. The solid blue and purple curves correspond to the best-fit MS and starburst (SB) FIR templates, respectively. The inset shows the posterior marginal likelihood of the FIR luminosity $\mathcal{L}(L_{\rm IR})$ obtained for the MS (blue) and starburst (purple) solutions. Here, we show a case where both the SB-type and MS-type templates provide reasonable fits, with a slight preference for the SB-type one.}
        \label{fig:sed}
\end{figure}

To robustly identify starburst galaxies, we focus on systems with direct detections in the FIR, where most of the star formation in dusty galaxies is re-emitted \citep[see e.g.][and references therein for a discussion about FIR-based starburst detections]{Rodighiero2011}. In heavily obscured systems, UV and optical light from young stars is efficiently absorbed by dust and reprocessed into thermal emission peaking at FIR wavelengths. As a result, SFRs derived from UV/optical SED fitting alone can be highly uncertain or systematically underestimated, especially for the most actively star-forming, dust-rich galaxies. We refer the reader to Sect.~4.2 in \citet{Malek2018}, Fig.~9 in \citet{li2024} and Appendix~C in \citet{accard2025} for discussions on the importance of FIR data for SFR and other galaxy quantities derivation from SED fitting with the \textsc{Cigale} \citep{cigale2019} and \textsc{Magphys} \citep{magphys2008} codes, respectively. To recover FIR fluxes in the \cosmos~field, we use the FIR fluxes from the \cosmosx~catalogue \citep{cosmosxid2024} developed within the \textit{Herschel} Extragalactic Legacy Project \citep[HELP;][]{Hurley_2016}. The HELP catalogues in the \cosmos~field combine data from the \textit{Spitzer}/MIPS $24\,\mu\rm m$ observations \citep{LeFloch2009}, the \textit{Herschel}/PACS Evolutionary Probe (PEP) survey \citep{Lutz2011}, and the \textit{Herschel}/SPIRE observations from the HerMES survey \citep{Oliver2012}. This catalogue provides probabilistic flux estimates for individual sources in \textit{Herschel}/PACS and SPIRE maps, which suffer from source confusion due to the large beam sizes at FIR wavelengths. Fluxes are extracted using the XID+ Bayesian deblending algorithm \citep{2017ascl.soft04012H}, which assigns fluxes to sources based on high-resolution prior positions. In \citet{cosmosxid2024}, the deblending is done progressively, first for MIPS data, then PACS data, and finally for SPIRE data, which improves the deblending performance. In this work, we adopt the version of \cosmosx~that uses the \cosmostwo~{\sc Farmer} positions as priors. This approach provides reliable estimates of FIR fluxes, which we will use to estimate reliable SFRs.

We first clean up our galaxy sample as follows. Galaxies without a known redshift are removed. Additionally, we select galaxies within the range $0.5 < z < 2$. As AGN have been identified and compiled in the XMM-\cosmos~catalogue \citep{2011ApJ...742...61S}, we remove them based on position matching, with a $.5''$ radius. Finally, to ensure good data quality and SED fitting, we only study galaxies with over 3 FIR ($\lambda > 20\,\rm \mu m$) band measurements with ${\rm S/N} \geqslant 3$. We are left with 4535 galaxies after this quality selection.

To derive $L_{\rm IR}$, we perform SED fitting using libraries of star-forming (both starbursts and MS) galaxy templates that have been calibrated and validated in the redshift and luminosity range of interest. Specifically, we adopt the templates of \citet{Magdis2012}, which builds on evolving dust emission templates capturing the variation of dust temperature and polycyclic aromatic hydrocarbon (PAH) features with redshift and galaxy properties, with starburst templates characterised by reduced PAH emission relative to the continuum, in agreement with observational studies \citep[e.g.][]{Elbaz2011}. In our analysis, for a given galaxy, the redshift is fixed, and the free parameters are the hardness of the radiation field $\langle U\rangle$ \citep[see][for details on this quantity]{Magdis2012} and the type of the template, either starburst-type or MS-type. Our fitting procedure involves scaling each template SED to the observed FIR photometry, accounting for redshift, and sampling the full posterior probability distribution of $L_{\rm IR}$, with the redshift fixed to the best photometric redshift from the catalogue. 

Figure~\ref{fig:sed} shows an example of the FIR SED fitting procedure for one galaxy in our sample, illustrating the comparison between the best-fit MS and starburst templates and the associated $L_{\rm IR}$ posterior distributions. In this case, both templates provide acceptable fits, with a mild preference for the starburst-type solution. To demonstrate that our fitting procedure behaves consistently when one template is clearly favoured, we also show in Appendix~\ref{appB} two representative examples (Figs.~\ref{fig:sed_ms} and~\ref{fig:sed_sb}) where the MS-type and starburst-type templates provide significantly better fits, respectively. In these cases, the preferred template also yields a correspondingly higher likelihood for $L_{\rm IR}$. Thus, whenever one template is statistically preferred over the other, the inferred $L_{\rm IR}$ follows unambiguously, and the choice of template does not introduce any ambiguity in $L_{\rm IR}$ or in the derived SFR.

In cases where the starburst and MS PDF modes are not clearly distinguished ($<0.8$ in template-based probability), the inferred $L_{\rm IR}$ values from both scenarios are comparable in amplitude, and do not impact our final selection based on SFR. For those cases, the typical difference is of the order of $10\%$, and it is thus not critical if the best-fit template is of the wrong type. Moreover, given our criterion in S/N and minimal number of measured fluxes, only a fraction of galaxies ($\sim5\%$) have acceptable SED fits with both MS-type and starburst-type templates. To distinguish starbursts from MS galaxies, we go beyond the template fitting procedure, and we compute for the best-fit template the corresponding SFR assuming the relation $\mathrm{SFR} = K \times L_{\rm IR}$, where $K = 1.09 \times 10^{-10}~\rm M_{\odot}\,\rm{yr}^{-1}\,\rm L_{\odot}^{-1}$ is the \citet{1998ARA&A..36..189K} conversion factor corrected to account for a \citet{2003PASP..115..763C} initial mass function. We then select galaxies with $\mathrm{SFR} > 4 \times \mathrm{SFR}_{\rm MS}$, with 4 being a common factor in the literature \citep[e.g.][]{Sargent2012}, where 
\begin{equation}
    \log\left(\frac{\rm SFR_{MS}}{\rm M_{\odot}\,yr^{-1}}\right) = m - m_{0} + a_{0}r - a_{1}(\max(0, m - m_{1} - a_{2}r))^2,
    \label{eq:sfr_ms}
\end{equation}
with $m_{0} = 0.5 \pm 0.07$,  $a_{0} = 1.5 \pm 0.15$, $a_{1} = 0.3 \pm 0.08$, $m_{1} = 0.36 \pm 0.3$, $a_{2} = 2.5 \pm 0.6$, $r = \log(1 + z)$, and $m = \log(M_{\star}/10^{9}\rm~M_{\odot})$, as defined in \citet{Schreiber2015}.

Finally, to have comparison samples, we perform mass-matching and redshift-matching for each starburst galaxy. We create samples of MS and quenched galaxies. These galaxies are selected using rest-frame NUV-r and r-J colours following a colour-colour criterion, where quenched galaxies satisfy $(\rm NUV - r) > 3 \times (r - J) + 1$ and $(\rm NUV - r) > 3$ as previously used in the literature, \citep[e.g.][]{williams2009, ilbert2013, laigle2016, laigle2018}. 
We quantify the quality of this matching by measuring the stellar mass difference between paired galaxies, finding a mean offset of $\sim2\times10^{-4}\,\rm dex$, a median consistent with zero, and a scatter of $\sim6\times10^{-3}\,\rm dex$.
We stress that FIR SED fitting is only used to identify dust-obscured high-SFR starbursts. MS and quenched galaxies do not require SED fitting, as their classification relies on rest-frame colours, with the additional condition of non-selection in the FIR starburst sample.

\section{Is there a redshift-dependent modulation of $\dfil$ in observations?}\label{sec:obs}

In this section, we explain how we perform cosmic web 2D reconstruction from the observational photometric data (Sect.~\ref{ssec:obs_cw}), and we test whether the relative distances of starburst, MS, and quenched galaxies to the cosmic web evolve with redshift in the \cosmos~field (Sect.~\ref{ssec:obs_res}). We quantify these trends while accounting for redshift uncertainties, providing a robust observational measurement of environment-dependent evolution.

\subsection{2D cosmic webs in the \cosmos~field}\label{ssec:obs_cw}

\begin{figure}[ht]
        \centering
        \includegraphics[width=\columnwidth]{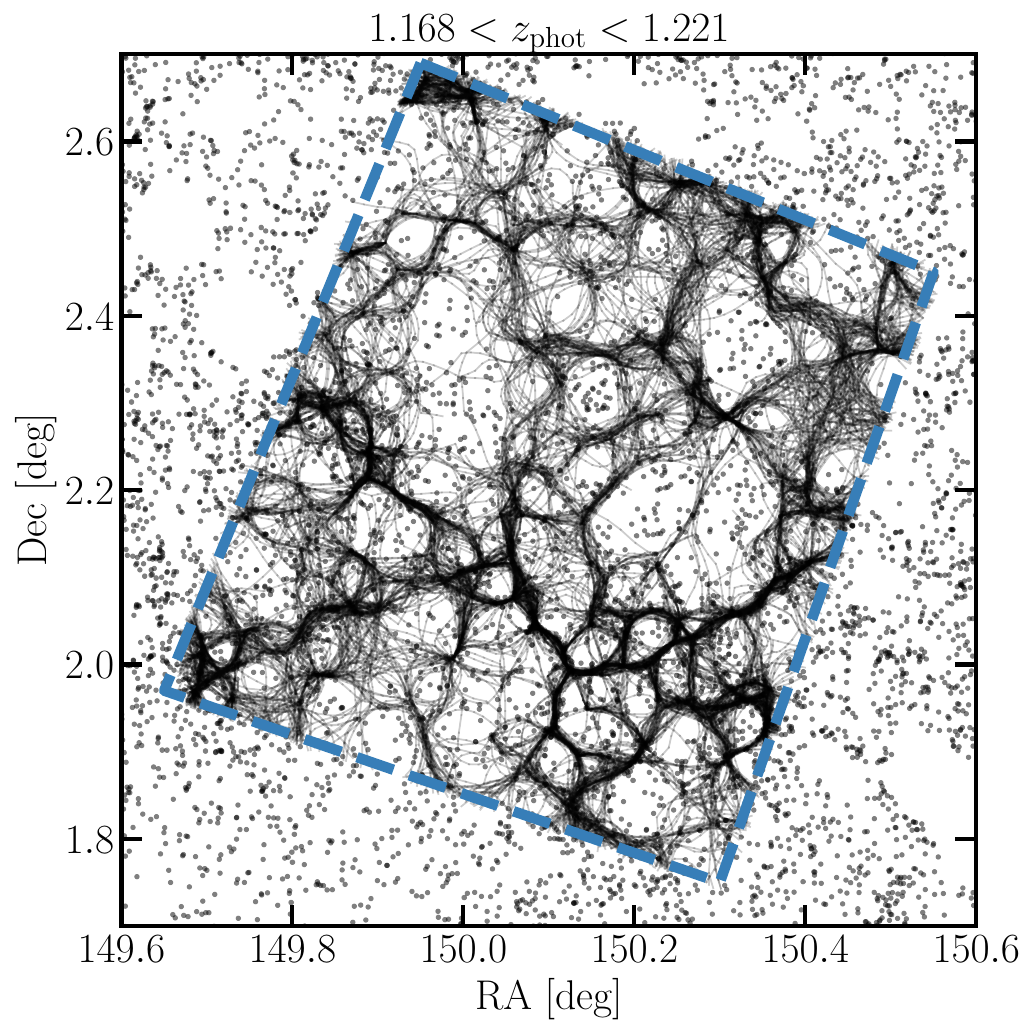}
        \caption{Galaxies in the \cosmos~field (grey dots) and 100 over-plotted realisations of the cosmic web for \cosmosweb~(filaments) in a slice centred around $z_{\rm phot}\simeq1.2$. The blue dashed contour represents the \cosmosweb~footprint.}
        \label{fig:cweb_slice}
\end{figure}

For observed galaxies, the cosmic web extraction method follows the one used by \cite{laigle2025}, using all galaxies in the \cosmos~field. We divide the galaxy sample into slices with a thickness chosen to match the photometric redshift uncertainties. To reduce redshift-dependent projection effects, we adopt a constant slice thickness in comoving length, corresponding to the median confusion length of the lowest-mass galaxies in the sample, i.e., the redshift width encompassing 68\% of the photometric redshift PDF for a typical galaxy. In \cosmostwo, galaxies more massive than $10^{9.5}\,\rm M_{\odot}$ have a confusion length smaller than $100\,h^{-1}\,\rm Mpc$ at $z_{\rm phot}<2$, and we therefore adopt this slice thickness for the reconstruction. In \cosmosweb, photometric redshifts are more precise, allowing a slice thickness of $80\,h^{-1}\,\rm Mpc$ up to $z_{\rm phot}=2$ for galaxies above $10^{9}\,\rm M_{\odot}$. Filaments are then identified with \disperse~using a $2.5\sigma$ persistence threshold, although results are stable down to $1.5\sigma$ \citep[see][for additional discussion on 2D density and cosmic web identification]{laigle2018}.

We compute 100 realisations of the cosmic web for the field, corresponding to samplings of the photometric redshift probability distributions. Figure~\ref{fig:cweb_slice} shows an example of a redshift slice in the \cosmos~field, illustrating the galaxy distribution and the corresponding cosmic web reconstructions in \cosmosweb~used in this analysis. 
The \cosmos~field contains several large masks, in particular due to the presence of bright stars in the foreground. Galaxies inside these masked regions are blended with the foreground sources or their photometry is polluted by the foreground, thus not in the catalogue. This leads to artificially empty regions in the 2D galaxy position distribution. Leaving these regions empty during the cosmic web extraction would distort the skeleton or lead to spurious filaments around the empty regions, since the edges of the masks would be locally artificially considered as high-density contrast regions. 

To circumvent this problem, in each tomographic slice, we generated randomly-distributed galaxy positions within the masks, with an average galaxy number density matching that of the tomographic slice. A different realisation of the mask filling is generated at each realisation of the skeleton described above. We provide a more quantitative comparison between the reconstructions with and without this correction in Appendix~\ref{appC}, where we show that near masked regions, distances to the filaments in the absence of correction can be underestimated. Not including mask correction in our analysis would therefore slightly bias the results.

With these corrections applied, we can now compute filament distances for each galaxy. For each galaxy, the 100 cosmic web realisations yield 100 $\mathrm{d}_{fil}$ estimates. We marginalise over redshift uncertainties by fully propagating the photometric redshift PDFs into the $\mathrm{d}_{fil}$ measurements. Each realisation is given a weight of 1/100 so that each galaxy contributes a total weight of unity. Uncertainties are consistently computed using these weights, ensuring that photometric redshift errors are properly propagated. We verified that alternative approaches such as adopting for each galaxy the mean or the median of $\log(\mathrm{d}_{fil}/\rm cMpc)$ over the 100 realisations and using the galaxy-to-galaxy scatter to estimate uncertainties, or excluding the 10\% highest and lowest $\mathrm{d}_{fil}$ values for each galaxy do not affect the results or the inferred uncertainties, indicating that the marginalised $\mathrm{d}_{fil}$ values are robust to outliers in the sampling. We also create a control data set by shuffling the distances, effectively attributing random $\mathrm{d}_{fil}$ values to the galaxies in the \cosmos~field.

\subsection{Evolution of the distance to cosmic filaments}\label{ssec:obs_res}

\begin{figure*}[ht]
        \centering
        \includegraphics[width=0.9\textwidth]{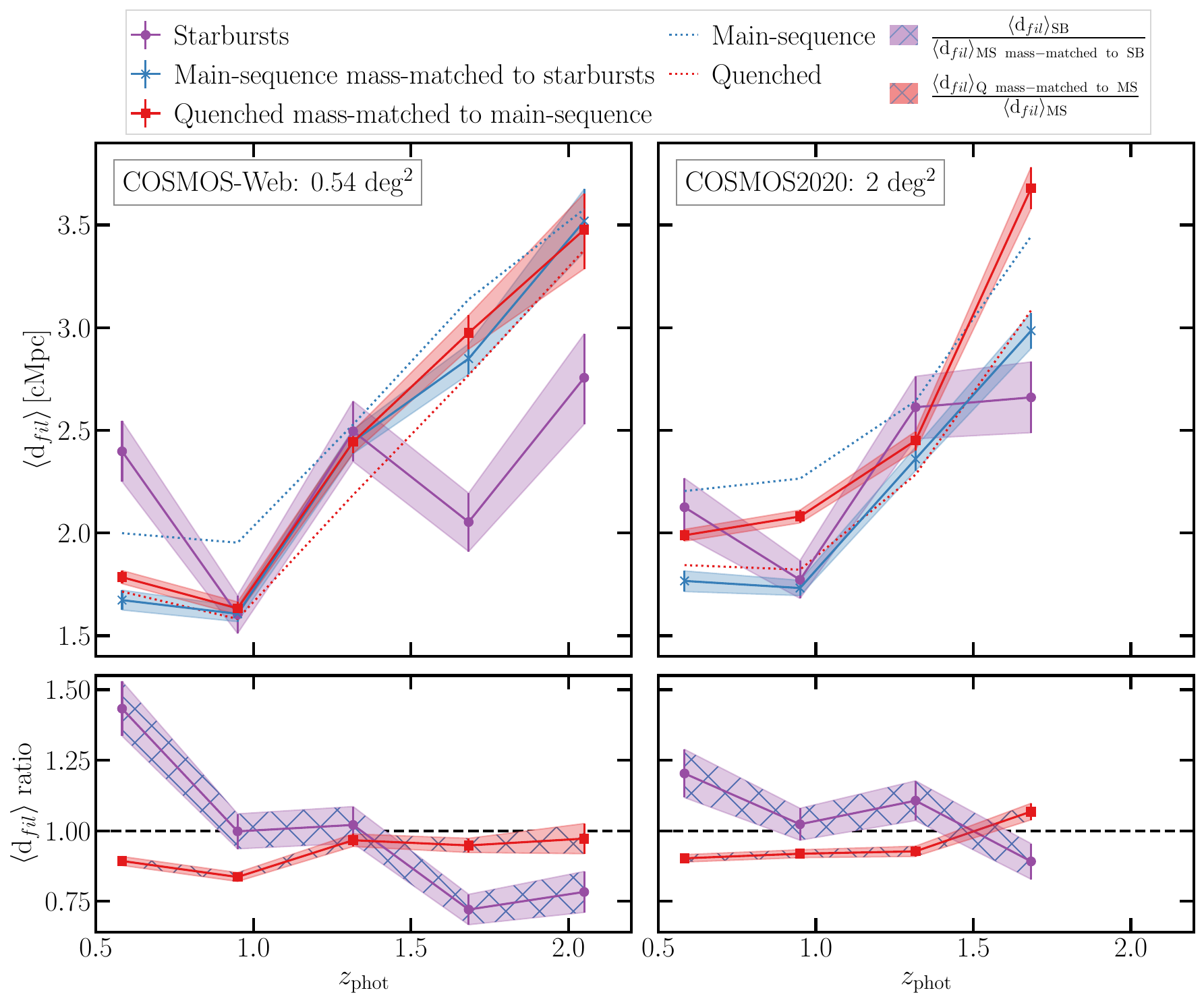}
        \caption{Upper row: Mean distance to the 2D cosmic web for starburst galaxies (purple solid line), MS galaxies mass-matched to starbursts (blue solid line), and quenched galaxies mass-matched to MS galaxies (red solid line) as a function of redshift in \cosmosweb~(left) and \cosmostwo~(right). The mean distances for the full MS and quenched samples are shown as blue and red dotted lines, respectively. 
        Lower row: Ratio of the mean distance of starbursts to that of mass-matched MS galaxies as a function of redshift (purple curve with blue hatching, derived from the purple and blue solid lines in the upper row). The red curve with blue hatching shows the ratio of the mean distance of quenched galaxies mass-matched to MS galaxies to that of the full MS sample (derived from the red solid line and blue dotted line in the upper row).
        The cosmic web is reconstructed independently for each dataset using the corresponding galaxy distribution. Distances to filaments are then measured with respect to the skeleton derived from the same catalogue.}
        \label{fig:d_z_total}
\end{figure*}

Figure~\ref{fig:d_z_total} presents the mean distance to the cosmic web for starburst, MS, and quenched galaxies in \cosmosweb~(left column) and \cosmostwo~(right column). To isolate environmental effects, the various populations are compared using stellar-mass-matched samples, including MS galaxies matched to starbursts and quenched galaxies matched to the main sequence. The redshift bins have a width of $\Delta z=0.3$, and each bin contains on average $\sim150$ starbursts in \cosmostwo~and $\sim50$ in \cosmosweb. 

The upper panels show that the relative location of the different galaxy populations with respect to the cosmic web evolves with redshift. At $z_{\rm phot} \gtrsim 1.25$, starburst galaxies are on average located closer to the cosmic web than mass-matched MS galaxies, while at lower redshifts they are found at larger distances. Quenched galaxies exhibit large mean distances at high redshift, comparable to those of the full MS population, but shift towards smaller distances for $z_{\rm phot} \lesssim 1.5$. 
The apparent differences in absolute distances to the cosmic web between \simba~and \cosmos~are mostly explained by the different galaxy number densities probed in the two datasets. In \simba, mean galaxy-galaxy separations correspond to higher 2D densities ($\Sigma \sim 3  {\rm~to~} 5 \,\rm gal/cMpc^{2}$), while in \cosmos~the lower sampling density yields $\Sigma \sim 0.2 {\rm~to~} 1 \,\rm gal/cMpc^{2}$ at similar redshifts. This motivates the use of distance ratios to compare relative environmental trends across datasets.

To better quantify these relative trends independently of the absolute distance scale, the lower panels show the ratios of the mean distance of starbursts to that of mass-matched MS galaxies, as well as the corresponding ratios for quenched galaxies. These ratios highlight the redshift at which the relative environmental behaviour of the different populations changes, and show qualitatively consistent evolutions in \cosmosweb~and \cosmostwo, within the statistical uncertainties. While differences in the absolute amplitude of the signal are observed between the two datasets, the redshift dependence of the relative trends is similar. \cosmosweb~provides improved redshifts, spatial resolution and goes to higher redshifts, whereas \cosmostwo~benefits from larger number statistics at lower redshift but lacks the highest-redshift bin probed by \cosmosweb.

To quantify the redshift evolution of the relative location of starburst galaxies with respect to the cosmic web, we focus on the ratios between the mean filament distances of starbursts and of mass-matched MS galaxies, shown in the lower panels of Fig.~\ref{fig:d_z_total}. Because the \cosmostwo~catalogue contains only four independent redshift bins for this measurement, we do not attempt to fit its redshift evolution independently, as such a small number of points does not allow a meaningful constraint on the slope. Instead, we perform quantitative fits using \cosmosweb~alone, and using a combined dataset including both \cosmosweb~and \cosmostwo~bins, which improves statistical robustness. In the combined dataset, we retain all galaxies from \cosmosweb~, and only include galaxies from \cosmostwo~that are not present in \cosmosweb, in order to avoid double-counting. Each galaxy is associated with the filament distances measured from the cosmic web reconstruction of its parent catalogue.
We stress that the functional forms adopted below are not meant to provide a physically motivated description, but rather to serve as a first-order parametric approach to test the null hypothesis of no evolution and to quantify the statistical significance of the observed trend. We consider two simple parametrisations, with a linear evolution with redshift,
\begin{equation}\label{eq:linfit}
    \frac{\dfil}{\dfil_{\rm MS}}(z) = a\,z + b ,
\end{equation}
and a power-law evolution,
\begin{equation}\label{eq:plfit}
    \frac{\dfil}{\dfil_{\rm MS}}(z) = A\,(1+z)^{\alpha} .
\end{equation}

In both cases, the parameters are constrained by maximising a Gaussian likelihood that explicitly accounts for the uncertainties on the measured ratios in each redshift bin. Parameter uncertainties are derived from the inverse Hessian matrix at the likelihood maximum, and the significance of the detected evolution is quantified by the ratio of the best-fitting slope (or exponent) to its uncertainty. We verified the validity of this approach by performing a full MCMC exploration of the parameter space for all fitted cases, using the same Gaussian likelihood and adopting broad, non-informative priors. In all cases, the posterior distributions are well described by multivariate Gaussians, with negligible parameter degeneracies and best-fitting values consistent with the maximum-likelihood estimates. The confidence intervals derived from the MCMC posteriors are fully consistent with those obtained from the inverse Hessian matrix at the likelihood maximum. Given the small number of data points and the near-Gaussian regime of the likelihood, we therefore adopt the Hessian-based uncertainties throughout, which provide an accurate estimation of the uncertainties.

For the starburst-MS ratio, \cosmosweb~alone already yields a clear detection of a redshift-dependent trend, whereas \cosmostwo~by itself does not provide sufficient leverage to identify such an evolution. The linear fit to the \cosmosweb~sample returns a negative slope $a = -0.39 \pm 0.07$, corresponding to a $6\sigma$ rejection of the null hypothesis $a=0$, with an intercept $b = 1.48 \pm 0.10$. The power-law fit leads to consistent conclusions, with an exponent $\alpha = -1 \pm 0.15$ ($6.6\sigma$ difference with $\alpha = 0$) and an amplitude $A = 2.13 \pm 0.26$. These results demonstrate a highly significant decrease in the relative distance of starbursts to the cosmic web with decreasing redshift.
Although \cosmostwo~alone is insufficient to detect a trend, combining \cosmosweb~with the \cosmostwo~measurements not already included in \cosmosweb~provides a useful consistency check, with the caveat that the filaments are not strictly the same between the two datasets. In this combined dataset, the constraints are slightly weakened compared to \cosmosweb~alone due to the shallower depth of \cosmostwo~and the reduced contrast in filament distance, but the overall trends remain robust. For SB/MS galaxies, the linear fit yields $a = -0.34 \pm 0.05$ ($6.7\sigma$) and $b = 1.41 \pm 0.07$, while the power-law fit gives $\alpha = -0.78 \pm 0.11$ ($7\sigma$) and $A = 1.81 \pm 0.16$. We note that adding information from \cosmostwo~always reduces the error bars on all fit parameters, but also reduces the slope. Overall, this results in a stronger exclusion of the null hypothesis. The same analysis for the distance ratio from \simba~(see Fig.~\ref{fig:d_z_prediction}) gives an exclusion of the null hypothesis at $\gtrsim 10 \sigma$ for both models, with $a = -0.18 \pm 0.015$ and $\alpha = -0.35 \pm 0.03$.
Nevertheless, the current analysis remains limited by the available starburst sample sizes, the different mass limits and amplitudes probed by the two catalogues, and the lack of high-redshift coverage in \cosmostwo. Larger and more homogeneous samples will be required to further refine the redshift dependence and fully characterise the amplitude of the observed trend.

Applying the same analysis to the quenched-to-MS ratios confirms that quenched galaxies tend to be located at progressively smaller relative distances from the cosmic web with decreasing redshift. In \cosmosweb~alone, the linear fit yields a positive slope $a = 0.072 \pm 0.02$ ($3.3\sigma$) with intercept $b = 0.82 \pm 0.02$, while the power-law fit gives $\alpha = 0.16 \pm 0.05$ ($3.2\sigma$) and amplitude $A = 0.8 \pm 0.03$. Using the combined \cosmosweb~+ \cosmostwo~dataset, the trend is respectively slightly less robust or similar with $a = 0.08 \pm 0.02$ ($4\sigma$) and $\alpha = 0.17 \pm 0.04$ ($4.1\sigma$). These results demonstrate a statistically significant increase in the mean distance of quenched galaxies relative to MS galaxies at higher redshift, in agreement with the qualitative evolution discussed above. The same analysis for the distance ratio from \simba~(see Fig.~\ref{fig:d_z_prediction}) gives an exclusion of the null hypothesis at $>10\sigma$ for both models, with $a = 0.18 \pm 0.005$ and $\alpha = 0.5 \pm 0.02$.

As a null test, we repeated the measurement shown in the upper-left panel of Fig.~\ref{fig:d_z_total} after randomly shuffling the $\mathrm{d}_{fil}$ values within each redshift bin. As expected, the signal vanishes as all populations lie on top of each other with only residual noise, with mean distances consistent across all galaxy types and centred around $\dfil \simeq 2.5\,\rm cMpc$. No systematic offset remains between starbursts, MS, and quiescent galaxies. This confirms that the observed trends in Fig.~\ref{fig:d_z_total} are not driven by random fluctuations. The shuffled realisation was generated and analysed in a blinded manner, in order to mitigate confirmation bias. In practice, the absence of any coherent signal in the shuffled case clearly distinguishes it from the real data.

For completeness, we also performed the same analysis using the distance to cosmic web nodes instead of filaments. As discussed in Appendix~\ref{appD}, the redshift evolution of the node-based distances (Fig.~\ref{fig:d_z_nodes}) shows a significantly reduced amplitude and does not allow firm conclusions to be drawn at this stage.

\section{Interpretation and validation}\label{sec:intp}

We now test the robustness of our observational trends (Sect.~\ref{ssec:intp_obs}), quantify projection effects in the cosmic web (Sect.~\ref{ssec:intp_2D3D}), and propose a toy model to describe the starburst-MS relative behaviours in \simba~(Sect.~\ref{ssec:intp_origin}). We then interpret the average distance to filament evolution for the different galaxy types (Sect.~\ref{ssec:intp_intp}). These steps consolidate the validity of our methodology and allow us to interpret the physical origin of the observed signal.

\subsection{Robustness of the observational catalogues}\label{ssec:intp_obs}

\begin{figure}[ht]
        \centering
        \includegraphics[width=\columnwidth]{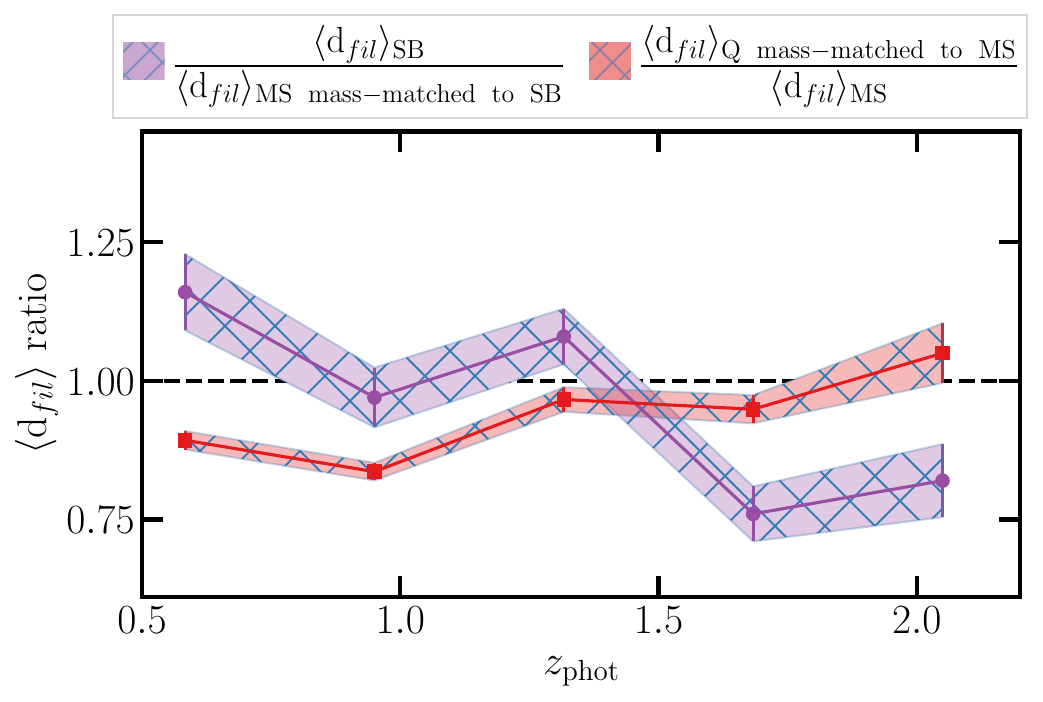}
        \caption{Same quantities as in the lower-left panel of Fig.~\ref{fig:d_z_total} after repeating our analysis on a starburst selection without using the MIPS-$24\,\rm \mu m$ band.}
        \label{fig:d_z_nomips}
\end{figure}

A key aspect of our analysis is to assess how sensitive our results are to the inclusion of the MIPS-$24\,\rm \mu m$ band, which plays a specific role in the template-based SED classification of starburst and MS galaxies. In our fitting scheme, the distinction between starburst and MS templates relies primarily on the mid-IR regime, where starbursts are characterised by a relative deficit of PAH emission compared to MS galaxies. As a consequence, removing the $24\,\rm \mu m$ band significantly reduces the ability of the SED fitting to discriminate between these two template families. We stress, however, that the choice of template family affects only the estimate of the total $L_{\rm IR}$ and does not affect the SFR-based definition of starbursts adopted in this work. To isolate this effect from more trivial signal-to-noise issues, we repeat the analysis by applying the same quality criterion as in the fiducial case, namely requiring at least three FIR bands detected at ${\rm S/N} > 3$, even after excluding the $24\,\rm \mu m$ data. Although this leads to a slightly different galaxy sample, the FIR luminosities remain reasonably well constrained by the remaining FIR measurements. In particular, the final starburst selection based on SFR, defined as $\rm SFR > 4\times SFR_{MS}$, is largely preserved, since it depends primarily on the total $L_{\rm IR}$ rather than on the detailed mid-IR spectral shape.

Under these conditions, we recover the same qualitative trends reported in Sect.~\ref{ssec:obs_res}, albeit with a somewhat reduced significance. We show the resulting distance ratios in Fig.~\ref{fig:d_z_nomips}. Here, $a=0$ and $\alpha=0$ (see models from Eqs.~\ref{eq:linfit} and~\ref{eq:plfit}) are excluded at $5.7\sigma$ and $6.3\sigma$ respectively for both the \cosmosweb~ and the \cosmosweb~+~\cosmostwo~datasets respectively. This weakening is expected, as removing the PAH-sensitive $24\,\rm \mu m$ band increases the uncertainty on the mid-IR part of the SED, which mildly propagates into the estimates of the total inferred $L_{\rm IR}$. Importantly, the persistence of the signal demonstrates that our main conclusions do not rely on a single photometric band. While the inclusion of $24\,\rm \mu m$ data improves the ability of the SED fitting to sample the mid-IR regime, the overall analysis remains stable, confirming that our results are driven by the global FIR properties of galaxies rather than by any individual wavelength.

We further validated that the measured separations between populations are not produced by residual systematics in the FIR luminosity fitting. We detail the procedure in Appendix~\ref{appE}. Figure~\ref{fig:d_z_random} shows the starburst/MS relative distances after randomising the FIR luminosities within redshift bins across our sample. By repeating the randomisation procedure 1000 times and averaging the resulting relative distances in each redshift bin, we ensure that the curves represent the statistical expectation and are not biased by any individual random realisation. In this control test, the resulting curve is consistent with a flat trend, and the error bars are much bigger at high redshift than for the non-randomised results due to large scattering in the $\mathrm{d}_{fil}$ values. This demonstrates that the signal identified in Sect.~\ref{ssec:obs_res} is not a trivial by-product of the fitting procedure or of the construction of our comparison samples, but arises from correlations between galaxy type and the cosmic web. Together, these tests establish the robustness of our classification and confirm that our main observational results are not artefacts of template choice, band coverage, or random fluctuations in $L_{\rm IR}$.

\subsection{Impact of projection effects on cosmic web reconstructions in \simba}\label{ssec:intp_2D3D}

\begin{figure}[ht]
        \centering
        \includegraphics[width=\columnwidth]{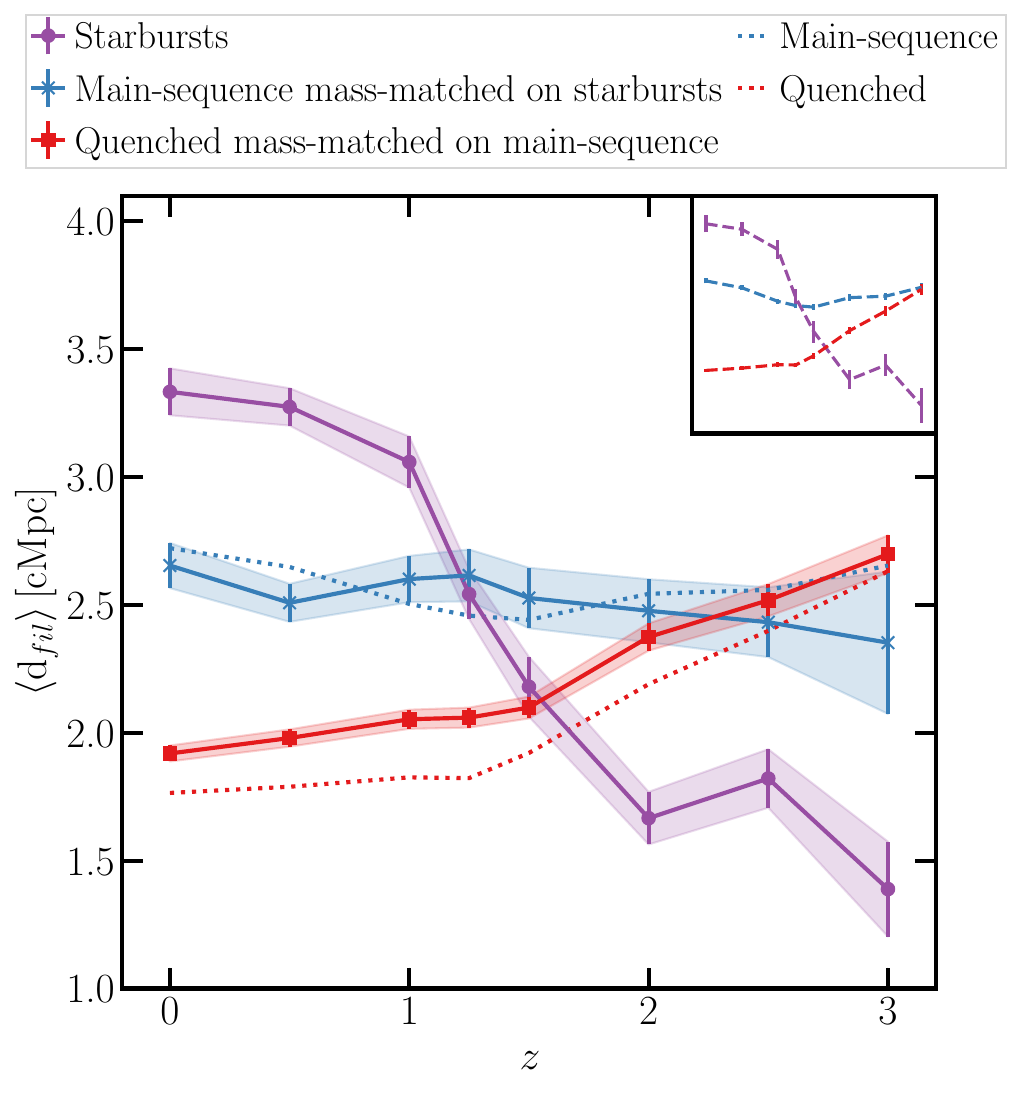}
        \caption{Mean distance to cosmic filaments measured in the full three-dimensional cosmic web for starburst (purple), MS (blue), and quenched (red) galaxies as a function of redshift in \simba, using the same selection and mass-matching procedure as in Fig.~\ref{fig:d_z_prediction}. This figure corresponds to the 3D counterpart of the upper-left panel of Fig.~\ref{fig:d_z_prediction}, without projection on a 2D plane. While the relative ordering between populations is preserved, the absolute distances are larger than in the projected case, and the redshift dependence of the trends differs, reflecting the impact of projection effects.}
        \label{fig:d_z_3d}
\end{figure}

Figure~\ref{fig:d_z_3d} shows the same measurement as the top left panel of Fig.~\ref{fig:d_z_prediction} performed using the full three-dimensional cosmic web in \simba, allowing us to quantify the impact of reconstructing the filament network in projections over a finite depth. Firstly, we note that the relative behaviour between starburst, MS, and quenched galaxies is stable between the projected and 3D measurements. In both cases, starburst galaxies are, on average, located farther from filaments than MS galaxies at $z\lesssim1$, while quenched systems remain the closest population for $z<2$. The corresponding distance ratios between populations are very similar to those obtained from measurements in projection.

However, in the 3D case, the redshift evolution of the mean distance to filaments differs significantly from that inferred in projection. Quenched galaxies exhibit a qualitatively similar behaviour, with decreasing distances toward low redshift. In contrast, MS galaxies show an approximately flat trend over the full redshift range, rather than the decreasing trend induced by projection effects. Starburst galaxies display a strong increase of $\dfil$ with decreasing redshift, crossing the MS curve around $z\simeq1.2$, without the flattening observed at $z \gtrsim 1$ in the projected case. This leads to a more pronounced separation from the main sequence at low redshift. Moreover, the absolute values of $\dfil$ are systematically higher than in the projected measurement. 

When reconstructing the cosmic web from galaxy positions projected on slices of depth $80\,\rm cMpc$, comparable to the size of the \simba~box ($100\,\rm cMpc$), distances to filaments are systematically reduced compared to the full three-dimensional case. This effect arises from several combined factors. First, the projection removes the line-of-sight component of the galaxy-filament separation, leading to a geometric contraction of distances. Second, the projection modifies the galaxy density field itself, as galaxies located at different radial positions are combined within the same slice which increases the density of the filamentary network and reduces the typical distance between galaxies and their nearest filament. Finally, the topology of the reconstructed cosmic web is altered: extended structures such as walls can be identified as filaments in projection, resulting in a denser and more space-filling filamentary network.
We also note that the intrinsic redshift dependence of the trends is distorted. In particular, projection induces a spurious decrease of $\dfil$ toward low redshift for MS galaxies and artificially flattens the evolution of starburst galaxies at $z \gtrsim 1$. These differences demonstrate that projection effects significantly affect both the amplitude and the redshift evolution of environmental trends, and that measurements based on fully three-dimensional cosmic web reconstructions provide qualitatively new information. This strongly motivates future observational efforts aimed at reconstructing the cosmic web in three dimensions, as will be enabled by upcoming wide and deep surveys.

\subsection{Evolution of the distance to cosmic filaments: a minimal environmental toy model for star-forming galaxies}\label{ssec:intp_origin}

\begin{figure}[ht]
        \centering
        \includegraphics[width=\columnwidth]{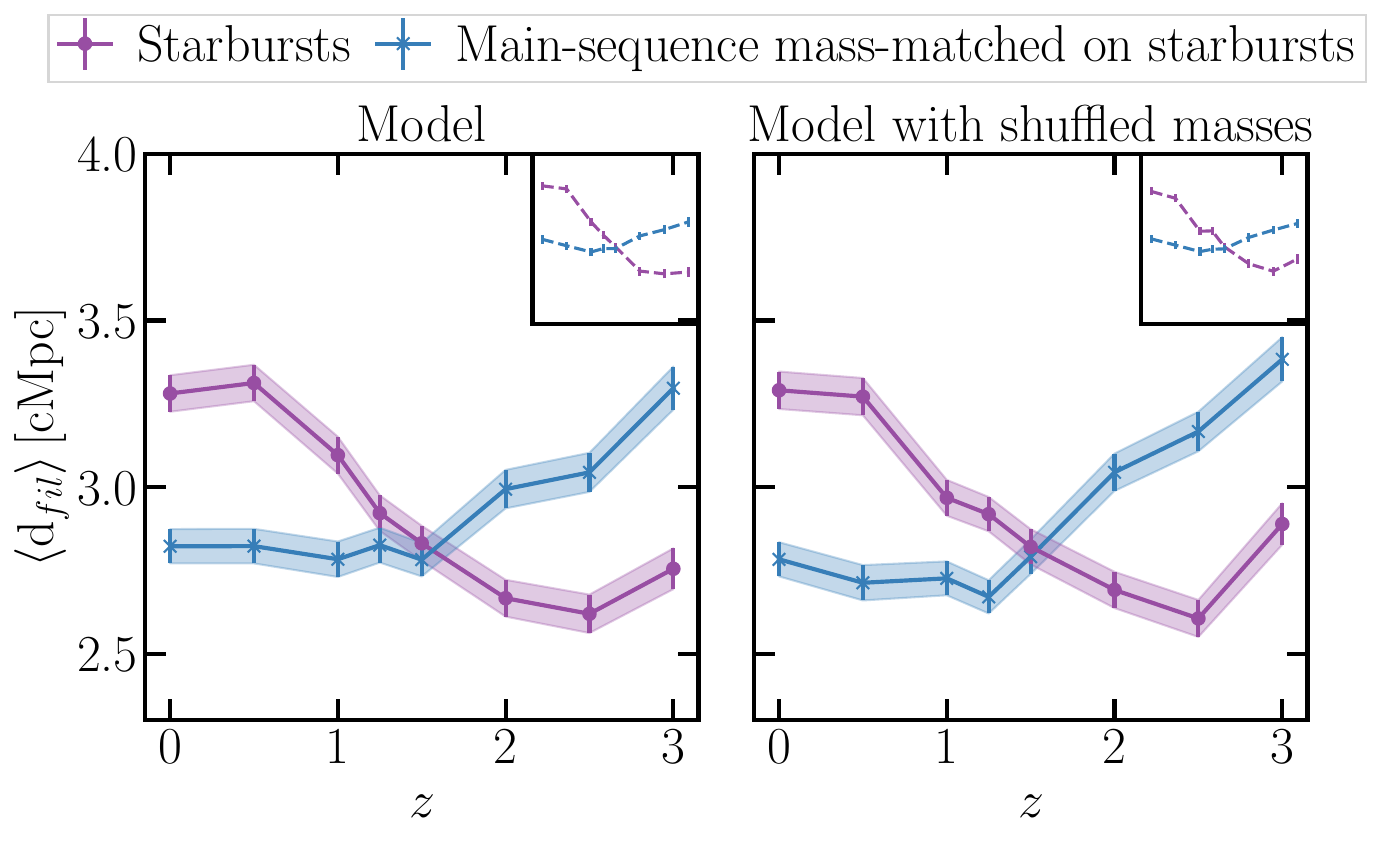}
        \caption{Left panel: Same quantities as in Fig.~\ref{fig:d_z_3d} after correction of the SFR modulation with distance and randomly picking a SFR deviation from a \citet{Schreiber2015} main sequence modified to fit \simba's main sequence. Here, starbursts have SFRs such as $\log(\rm SFR/\rm SFR_{MS}) > 0.4\, dex$. Right panel: Same quantities as in the left panel with shuffled galaxy masses. Although the $\dfil$ values change, the lines cross as in Fig.~\ref{fig:d_z_3d}, and the general trends are similar. This allows us to test the effect of modulation of SFR with the distance to filaments characterised in \citetalias{jego2026}.}
        \label{fig:d_z_model}
\end{figure}

To investigate whether the modulation of star formation activity with distance to cosmic filaments, characterised in the \simba~simulation by \citetalias{jego2026}, can account for the separation observed between MS and starburst galaxies, we construct a complementary synthetic sample of SFG. The purpose of this approach is to explicitly include the environment-dependent modulation of SFR found for SFG in \simba~in a controlled framework and to test whether it can reproduce the trends seen in the simulation. Importantly, no additional correlation between galaxy properties and environment is introduced beyond the sSFR-$\dfil$ modulation measured in \citetalias{jego2026}.

We first define the star-forming main sequence using the redshift-dependent analytic form of \citet{Schreiber2015} (Eq.~\ref{eq:sfr_ms}). This functional form is refitted to the \simba~star-forming population at each snapshot, so that the reference main sequence reflects the intrinsic normalisation and evolution of the simulation rather than the observational calibration. Using the same \simba~data, we then quantify the modulation of the specific star formation rate with distance to filaments by measuring, at fixed stellar mass and redshift, the variation of sSFR as a function of $\mathrm{d}_{fil}$. We derive a modulation $\Delta \log({\rm sSFR})$ in bins of distance to filaments, defined as the offset of the average sSFR relative to the median MS predicted by \simba~at fixed stellar mass. This modulation, measured in \citetalias{jego2026}, captures the first-order dependence of star formation activity on cosmic web environment in the simulation.

For each galaxy, we generate a new specific star formation rate by sampling a double-Gaussian distribution in $\log({\rm sSFR}/{\rm sSFR}_{\rm MS})$ (see Fig.~19 in \citealt{Schreiber2015}). The first component represents the MS population, while the second accounts for a starburst tail at higher sSFR. The relative weight of the starburst component evolves with redshift following a starburst fraction $f_{\rm SB}(z)~=~0.012 \times\min(z,1)$, motivated by the empirical evolution reported in \citet{Sargent2012} and \citet{Bethermin2012}. The environmental dependence is included by applying the corresponding $\Delta \log({\rm sSFR})$ as an additional offset to the sampled distribution. Galaxies are then classified as starbursts if their sampled star formation rate satisfies $\log({\rm SFR}/{\rm SFR}_{\rm MS}) > 0.4\,\rm dex$, although similar results are obtained for thresholds of $0.2$ or $0.6\,\rm dex$. To isolate the role of stellar mass, we repeat the same procedure after randomly shuffling stellar masses among galaxies at each redshift, while keeping their distances to filaments fixed. This removes any correlation between mass and environment, allowing us to test whether the resulting trends are driven purely by the imposed sSFR modulation.

The resulting trends, shown in Fig.~\ref{fig:d_z_model}, closely reproduce the qualitative behaviour observed directly in \simba, with starburst galaxies found at smaller distances to filaments than MS galaxies at high redshift, and at higher distances at low redshift, with an intersection around $z\sim1.5$. This indicates that the modulation of sSFR with distance to the cosmic web characterised in \citetalias{jego2026} can account for the dominant contribution to the observed separation, with $\dfil$ amplitudes remaining similar at low redshift. Although the modelled behaviours of $\dfil$ for the MS and starburst galaxies do not exactly match those seen in \simba~at high redshift (the median relations shown in the top right corners are a closer match), this analysis hints that the zeroth-order signal arises from the environment-dependent modulation of star formation activity. Shuffling masses have a negligible effect (see Fig.~\ref{fig:d_z_model}), confirming that the relationship between filament distance and SFR exists independently of the underlying link between environment and mass.

This exercise does not aim at providing a physical explanation for the modulation itself, but demonstrates that the sSFR-filament distance relation measured and physically discussed in \citetalias{jego2026} is able to qualitatively reproduce the observed differential evolution of $\dfil$ between MS and starburst galaxies.

\subsection{Redshift-dependent modulation of star formation across the cosmic web}\label{ssec:intp_intp}

The trends observed in \cosmos~can be further discussed in the context of the \simba~analysis presented in \citetalias{jego2026}, where the dependence of star-forming activity on distance to cosmic filaments was quantified after controlling for stellar mass. In that work, the modulation of the sSFR residuals ($\Delta \log({\rm sSFR})$) with filament distance is detected for the overall SFG population at all explored redshifts ($z\leqslant 3$), indicating that the large-scale environment is associated with star-formation properties beyond the primary dependence on stellar mass (see details in Appendix~\ref{appF} and Fig.~\ref{fig:DsSFR_dfil}). \citetalias{jego2026} further shows that the origin of this modulation varies with redshift. At high redshifts ($z\sim2-3$), the majority of SFG are centrals, and the enhanced sSFR observed closer to filaments is consistent with efficient gas supply along the cosmic web. In this regime, the large-scale structure may primarily act by facilitating the availability of gas feeding star formation. At lower redshift ($z\sim0$), the SFG population becomes increasingly dominated by satellite galaxies. For these systems, the observed dependence of sSFR on filament distance is likely mediated by the local environment of satellites, which itself reflects the underlying large-scale structure, with SFR suppression at intermediate distances and a regain of star-formation close to filaments due to a balance between gas heating and accretion.

The \cosmos~measurements are consistent with this picture. In particular, the behaviour of the starburst population appears to evolve differently from that of MS galaxies with redshift. At higher redshifts, starburst galaxies are found on average closer to filaments, which may simply reflect their presence in regions where gas reservoirs are more readily available. Toward lower redshift, starburst galaxies populate environments, on average, further away from the cosmic filaments than MS galaxies, which could indicate that internal baryonic processes, including feedback or stochastic gas inflows, increasingly prevent the occurrence of starburst episodes in the cosmic web. For quiescent galaxies, our analysis further shows that they tend to reside closer to filaments, a trend that strengthens toward lower redshift, in parallel with the well-known increase of their overall fraction. Several mechanisms may contribute to this behaviour, including environmental quenching in dense regions or the preferential shutdown of star formation in massive systems going from MS (or starburst) to quenched in situ.

A key aspect of our interpretation concerns the different physical timescales involved in starburst activity and large-scale environmental processes. Starbursts are intrinsically short-lived events ($\sim 10-100\,\rm Myr$), whereas the influence of the cosmic web on galaxy growth operates over much longer timescales ($\gtrsim 1\, \rm Gyr$). As a result, any connection between starburst episodes and filamentary environments should not be interpreted as a direct causal link on individual objects. Instead, the observed trends reflect statistical correlations between the likelihood of experiencing a starburst phase and the long-term environmental history of galaxies.
In this context, starbursts should be regarded as transient tracers of underlying environmental dependencies of star formation, rather than direct probes of instantaneous environmental triggering. This interpretation is consistent with the absence of a strong merger-driven contribution in \simba~for the studied redshift range reported in \citetalias{jego2026}, as well as with our toy model (Sect.~\ref{ssec:intp_origin}), which shows that the observed differences in filament distance can emerge from a redshift-dependent modulation of the star-forming main sequence with environment, without requiring a direct one-to-one correspondence between starburst events and local triggers such as mergers.

\section{Conclusion}\label{sec:ccl}

In this work, we investigate the link between galaxy types, which are distinguished as starburst, MS, and quenched, and their locus with respect to the cosmic web filaments. The cosmic web is reconstructed from galaxy positions using \disperse. 

We first use the \simba~cosmological hydrodynamical simulation to predict the redshift evolution of the mean distance to filaments for starburst, MS, and quenched galaxies, after controlling for stellar-mass dependencies. These predictions are then confronted with measurements in the \cosmos~field, where the multi-wavelength photometric coverage yields highly accurate photometric redshifts, enabling a reliable reconstruction of the cosmic web. In the \cosmos~field, starbursts are robustly identified through FIR SED fitting, which allows to recover FIR luminosities and accurate SFRs. Finally, we assess the robustness of the observational signal and explore a simple toy model that can partly account for the observed trends.

Our main results can be summarised as follows:

\begin{itemize}
    \item In \simba, we find a clear differential redshift evolution of the average galaxy populations. Starburst galaxies are preferentially located closer to filaments at high redshift ($z \geqslant 1$), while at low redshift ($z \leqslant 1$), they are found at larger distances from filaments. Quenched galaxies are increasingly associated with smaller distances from filaments toward low redshift, whereas MS galaxies consistently occupy intermediate environments. The starburst and MS trends intersect around $z \sim 1$. These results are robust to moderate variations in the definition of starburst galaxies. The differences in average distance to filaments between populations are consistent with redshift-dependent modulations of sSFR by the distance to filaments studied by \citet[][\citetalias{jego2026} in this study]{jego2026} in the \simba~simulation.
    \item In COSMOS, the relative behaviour of populations is consistent with simulations: in \cosmosweb~we detect a significant ($\gtrsim 6\sigma$) starburst-MS separation decreasing with redshift, while \cosmostwo~yields weaker constraints.
    \item Projection effects (2D vs. 3D cosmic web reconstructions) reduce the signal amplitude and affect both slope and redshift dependence. In 2D, the mean distance to filaments of the MS population decreases toward low redshift, with a stronger decrease for quenched galaxies, while starbursts show an overall flat redshift trend. In 3D, starbursts are found at progressively larger filament distances at lower redshift, MS galaxies show little evolution, and quenched systems exhibit a systematic decrease from $z=3$ to $z=0$.
    \item Robustness tests (band exclusion, shuffled luminosities, alternative web reconstructions) confirm that the observed separation is not driven by random noise or fitting artefacts.
\end{itemize}

Overall, our results provide initial evidence that the environmental imprint predicted by cosmological simulations is already within reach of current deep multi-wavelength photometric surveys. While the amplitude of the signal remains affected by projection and statistical limitations, the consistent ordering of galaxy populations in both data and simulations suggests a coherent environmental imprint of the cosmic web on galaxy star formation properties. Future wide and deep surveys with \textit{Euclid}, JWST, PFS, and \textit{Roman} will allow 3D reconstructions over large volumes, offering the opportunity to confirm these trends with much higher precision and to disentangle the interplay between large-scale environment, local density and anisotropy, and feedback processes in driving galaxy evolution.

\begin{acknowledgements}
      The authors thank the referee for insightful comments and suggestions that helped to improve the presentation of the paper.
      B. J. is supported by a CDSN doctoral studentship through the ENS Paris-Saclay. 
      L. W. acknowledges funding from the project ‘Clash of the titans: deciphering the enigmatic role of cosmic collisions’ (with project number VI.Vidi.193.113) of the research programme Vidi which is (partly) financed by the Dutch Research Council (NWO).
      French COSMOS team members are partly supported by the Centre National d’Etudes Spatiales (CNES). OI acknowledges the funding of the French Agence Nationale de la Recherche for the project iMAGE (grant ANR-22-CE31-0007).
      We acknowledge the High-Performance Computing Center of the Astronomical Observatory of Strasbourg for supporting this work by providing scientific support and access to computing resources.
      We thank Romeel Davé for making the \simba~simulation suite publicly available.
      We made extensive use of the {\tt numpy} \citep{oliphant2006guide, van2011numpy}, {\tt scipy} \citep{2020SciPy-NMeth}, {\tt astropy} \citep{1307.6212, 1801.02634}, {\tt caesar} \citep{2011ApJS..192....9T}, and {\tt matplotlib} \citep{Hunter:2007} Python packages.\\
      The data used and produced in this work are available upon reasonable request to the corresponding author.
\end{acknowledgements}

\bibliographystyle{aa}
\bibliography{main}

\begin{appendix}

\onecolumn
\section{Galaxy selections in \simba}\label{appA}

\begin{table}[ht]
    \caption{Number of galaxies per type, per redshift, and per SFG selection.}
    \centering
    \def\arraystretch{1.4}
    \begin{tabular}{|c|cccccccc|}
      \hline
      $z$ & $0$ & $0.5$ & $1$ & $1.25$ & $1.5$ & $2$ & $2.5$ & $3$ \\
      \hline
      \hline
      \multicolumn{9}{|c|}{$\log(\mathrm{SFR}_{\rm SB}/\mathrm{SFR}_{\rm MS}) > 0.2\,\rm dex$}\\
      \multicolumn{9}{|c|}{$\log(\tau_{\rm SB}/\tau_{\rm MS}) < -0.2\,\rm dex$}\\
      \hline
      Q & 27628 & 17526 & 12057 & 10444 & 9384 & 9933 & 7037 & 5787 \\
      MS & 23868 & 22620 & 24498 & 24469 & 24205 & 19695 & 19009 & 18534 \\
      SB & 2166 & 3589 & 1473 & 1356 & 810 & 898 & 985 & 225 \\
      \hline
      \hline
      \multicolumn{9}{|c|}{$\log(\mathrm{SFR}_{\rm SB}/\mathrm{SFR}_{\rm MS}) > 0.4\,\rm dex$}\\
      \hline
      MS & 25459 & 24938 & 25602 & 25435 & 24780 & 20331 & 19635 & 18678 \\
      SB & 575 & 1271 & 369 & 390 & 235 & 262 & 359 & 81 \\
      \hline
      \hline
      \multicolumn{9}{|c|}{$\log(\mathrm{SFR}_{\rm SB}/\mathrm{SFR}_{\rm MS}) > 0.2\,\rm dex$}\\
      \hline
      MS & 22336 & 20796 & 23859 & 23941 & 23852 & 19526 & 18779 & 18469 \\
      SB & 3698 & 5413 & 2112 & 1884 & 1163 & 1067 & 1215 & 290 \\
      \hline
      \hline
      \multicolumn{9}{|c|}{$\log(\tau_{\rm SB}/\tau_{\rm MS}) < -0.4\,\rm dex$}\\
      \hline
      MS & 23025 & 21494 & 22456 & 22677 & 22333 & 18746 & 18595 & 17745 \\
      SB & 3009 & 4715 & 3515 & 3148 & 2682 & 1847 & 1399 & 1014 \\
      \hline
    \end{tabular}
    \tablefoot{Number of galaxies per type (Q=Quenched, MS=Main-Sequence, SB=Starburst) in each \simba~snapshot for the four selections proposed in Sect.~\ref{ssec:simba_evol}. The number of quenched galaxies does not change between selections.}
    \label{tab:ngals}
\end{table}
\twocolumn
Table~\ref{tab:ngals} reports the number of galaxies in each population (quenched, MS, and starburst) for all snapshots and selection criteria used in this work. These numbers correspond to the samples used in Figs.~\ref{fig:d_z_prediction} and \ref{fig:d_z_prediction_ratios}. The number of galaxies in each population shows the expected hierarchical structure of the selection. MS galaxies dominate the sample at high redshifts, while quenched galaxies represent a large but decreasing fraction toward higher redshift. Starburst galaxies remain a sub-dominant population in all cases, typically representing a few percent of the total galaxy sample, with a mild increase depending on the adopted selection criterion.

The evolution with redshift reflects both the growth of the star-forming population at early times and the progressive build-up of the quenched population at late times. Importantly, even in the highest-redshift bins, all selections retain sufficient statistics (typically several hundred to a few thousand galaxies per population), ensuring that the measured mean filament distances and ratios are not dominated by small-number noise. We also note that the relative proportions between starburst and main-sequence galaxies remain broadly stable across redshift for most selection criteria, with starbursts consistently representing $\sim 5-15\%$ of the star-forming population. This stability supports the robustness of the inferred redshift evolution in Fig.~\ref{fig:d_z_prediction}.

\section{Examples of FIR SED fitting}\label{appB}

\begin{figure}[ht]
        \centering
        \includegraphics[width=\columnwidth]{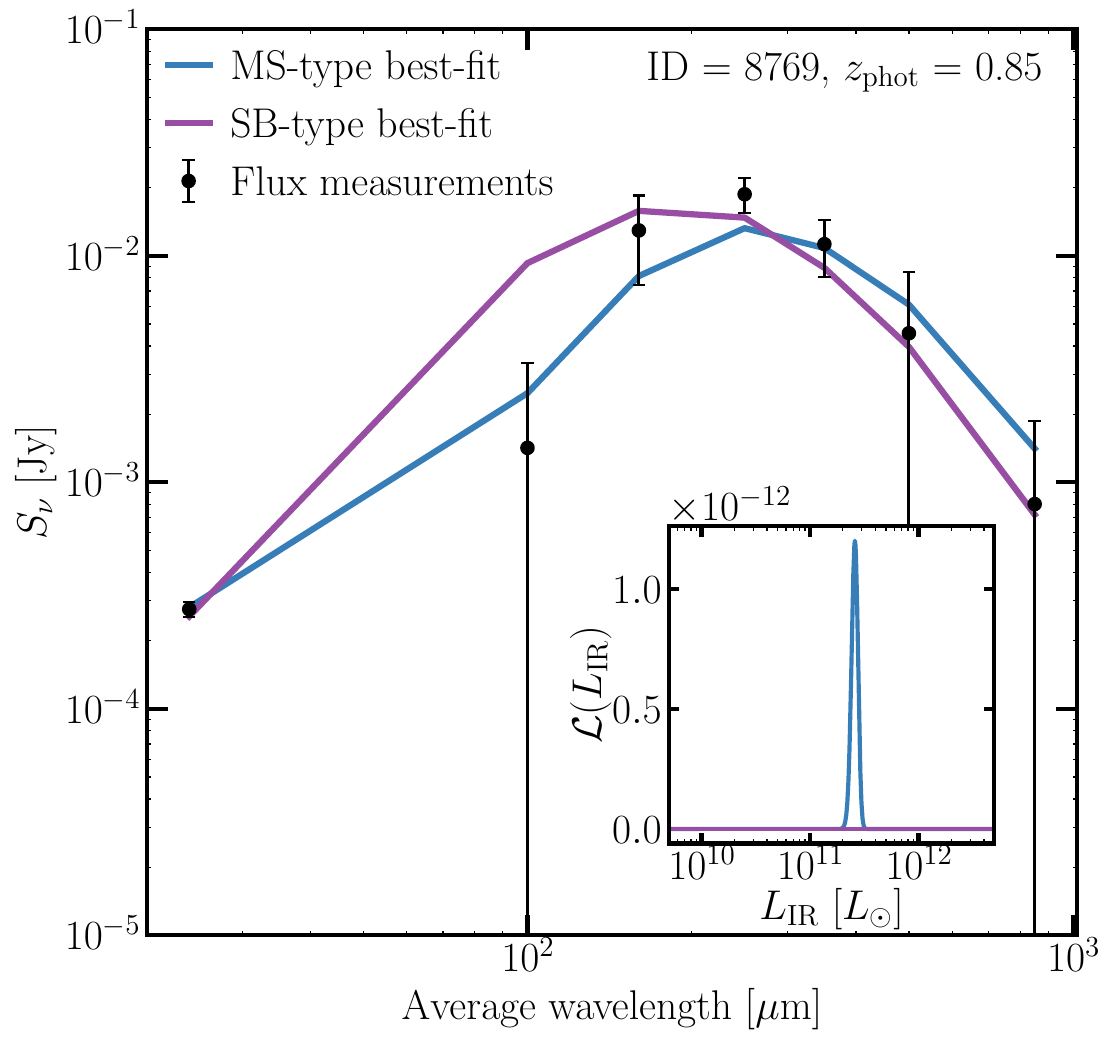}
        \caption{Same quantities as in Fig.~\ref{fig:sed} with the MS template being a better fit.}
        \label{fig:sed_ms}
\end{figure}

\begin{figure}[ht]
        \centering
        \includegraphics[width=\columnwidth]{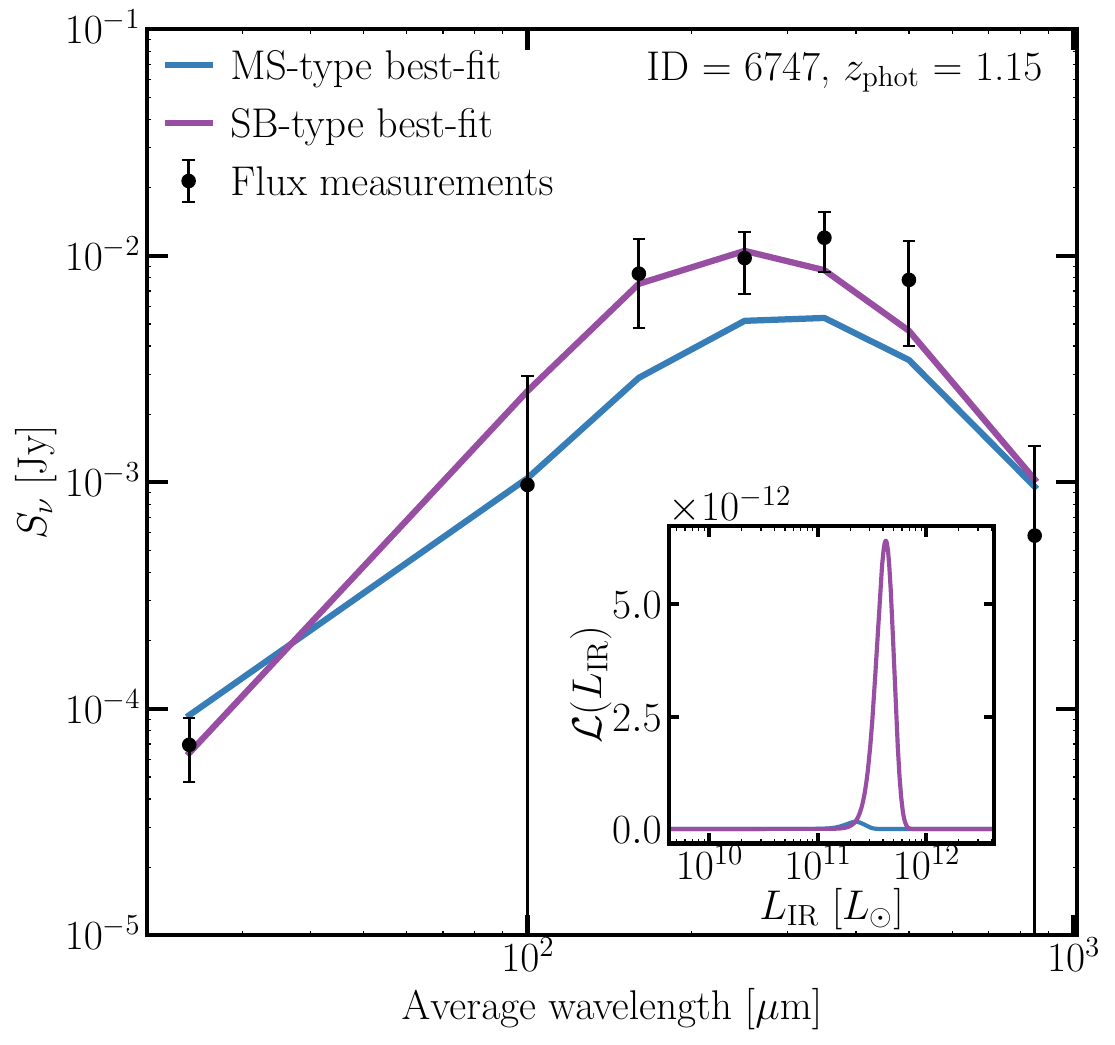}
        \caption{Same quantities as in Fig.~\ref{fig:sed} with the starburst template being a better fit.}
        \label{fig:sed_sb}
\end{figure}

Figures~\ref{fig:sed_ms} and \ref{fig:sed_sb} show two representative examples of FIR SED fitting for galaxies in our \cosmos~sample as described in Sect.~\ref{ssec:sample_sel}, where the MS and starburst templates provide a significantly better fit, respectively. In both cases, the preferred template is clearly favoured by the photometry and yields a higher likelihood for the FIR luminosity $L_{\rm IR}$. The corresponding posterior distributions show little overlap, indicating that the inferred $L_{\rm IR}$ is robust and unambiguous. These examples illustrate that, when the FIR SED shape is well constrained by the data, the resulting $L_{\rm IR}$ (and thus SFR) estimates are robust. We stress that the MS-type or SB-type nature of the best-fitting template is not used for our starburst classification, which relies solely on the offset from the star-forming main sequence, but these examples validate the reliability and precision of the inferred $L_{\rm IR}$.

\section{Impact of photometric masks on cosmic web reconstructions}\label{appC}

\begin{figure}[ht]
    \centering
    \includegraphics[width=\columnwidth]{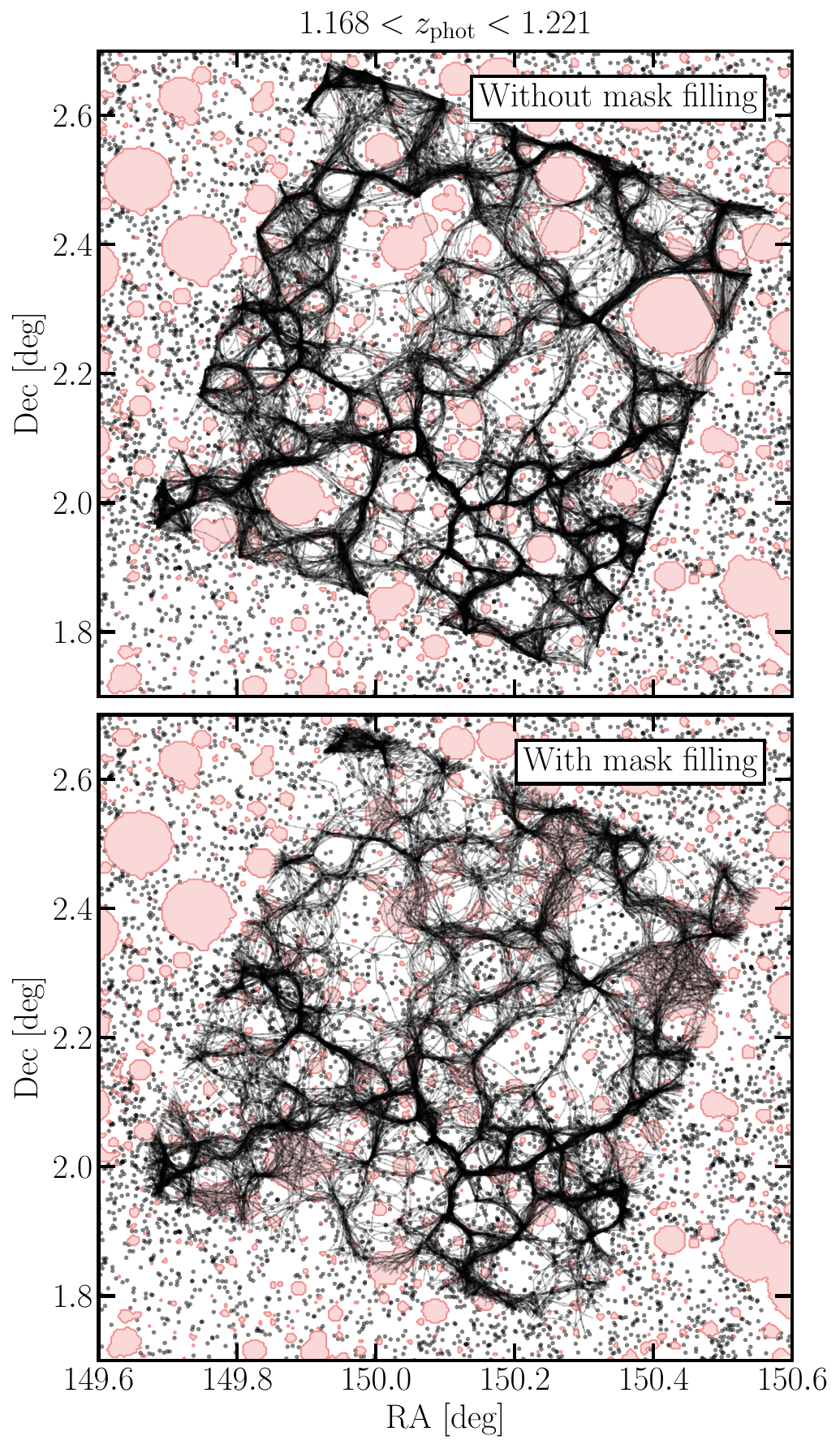}
    \caption{
    Galaxies in the \cosmos~field (grey dots) and 100 over-plotted realisations of the cosmic web in a redshift slice centred at $z_{\rm phot}\simeq1.2$. Upper panel: \disperse~reconstruction performed directly on the masked galaxy distribution. 
    The survey masks are visible as wide red regions completely devoid of galaxies, which induce artificial filamentary structures wrapping around their boundaries. Lower panel: Reconstruction after random in-painting of the masked regions prior to running \disperse~(method adopted in this work), which mitigates these edge-induced artefacts.}
    \label{fig:cweb_slice_compar}
\end{figure}

\begin{figure}[ht]
    \centering
    \includegraphics[width=\columnwidth]{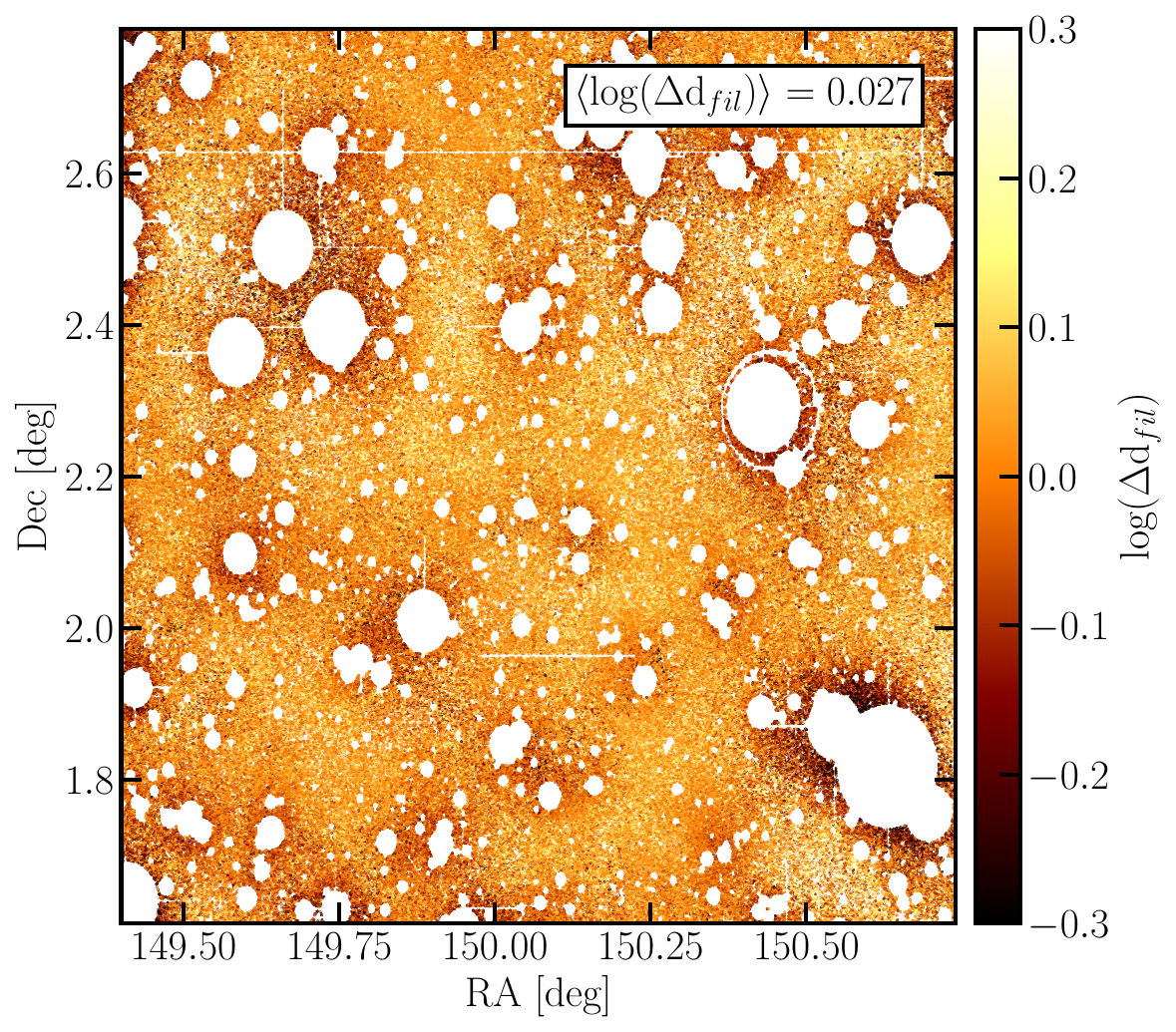}
    \caption{
    Logarithmic ratio of filament distances, $\log(\Delta {\rm d}_{fil}) = \log({\rm d}_{\rm filled} / {\rm d}_{\rm unfilled})$, for all galaxies in the \cosmos~field. Here, ${\rm d}_{\rm filled}$ is the distance to the nearest filament measured after in-painting the survey masks, while ${\rm d}_{\rm unfilled}$ is measured without correcting for masked regions. Negative values occur near masked areas, where artificial gaps in the galaxy distribution lead to underestimated distances, whereas positive values appear farther from masks. For visualisation purposes, the colour bar is clipped to $[-0.3, 0.3]$, but the actual values can exceed this range. The mean of this ratio across the field is indicated in the top right (0.027). All redshift slices are stacked, taking the average filament distance per galaxy for each measurement method.}
    \label{fig:dfil_compar}
\end{figure}

\begin{figure}[ht]
    \centering
    \includegraphics[width=\columnwidth]{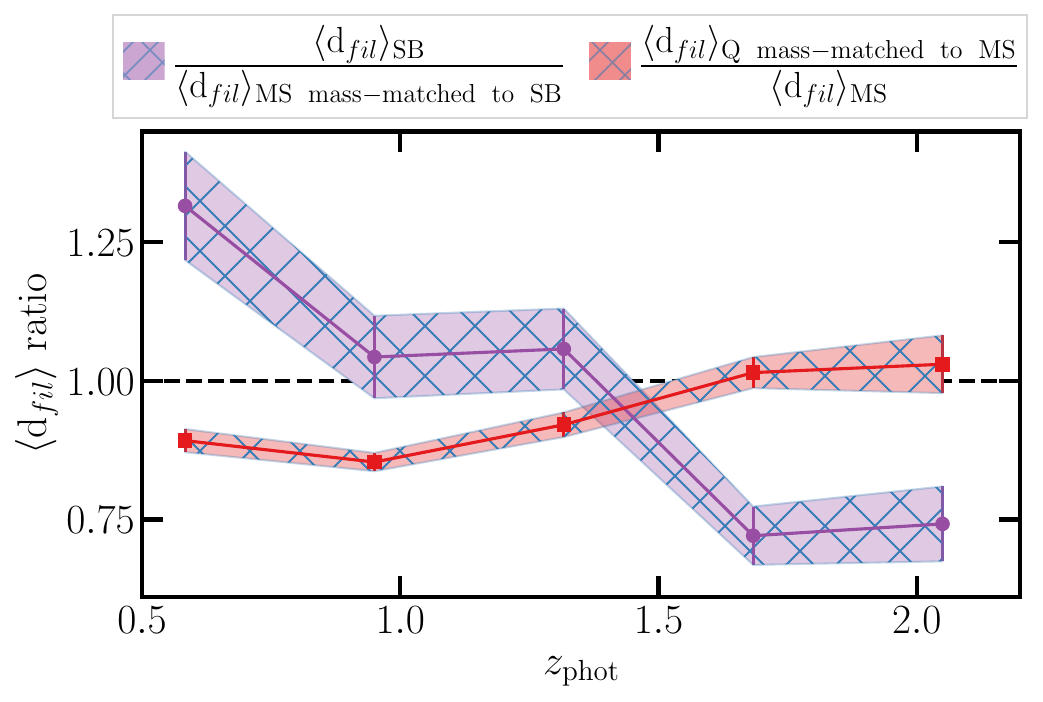}
    \caption{Same quantities as in the lower-left panel of Fig.~\ref{fig:d_z_total}, but using the cosmic web reconstructions without correcting for the masked regions.}
    \label{fig:d_z_compar}
\end{figure}

The presence of survey masks in the \cosmos~field, mainly due to bright foreground stars, creates empty regions in the galaxy spatial distribution. Leaving these masked regions uncorrected in the cosmic web reconstruction can generate spurious filaments around the edges, as the boundaries are locally interpreted as high-density contrasts. 

Figures~\ref{fig:cweb_slice_compar} and~\ref{fig:dfil_compar} illustrate the effect of these masks on 2D cosmic web extractions in a representative redshift slice at $z_{\rm phot}\simeq1.2$. Figure~\ref{fig:cweb_slice_compar} compares the filamentary skeleton obtained directly on the masked galaxy distribution (upper panel) to the skeleton after random in-painting of the masked regions (lower panel). The in-painting method mitigates edge-induced artefacts and restores a more realistic reconstruction. 

Figure~\ref{fig:dfil_compar} quantifies the impact on filament distances, showing the logarithmic ratio $\log({\rm d}_{\rm filled}/{\rm d}_{\rm unfilled})$, for all galaxies. ${\rm d}_{\rm filled}$ and ${\rm d}_{\rm unfilled}$ are computed with and without filling the masks respectively. Near masked regions, distances to filaments are systematically underestimated when masks are not corrected, whereas farther from masks, the effect is reversed. The mean difference across the field is small (0.027 in $\log({\rm d}_{fil})$), indicating that the global impact of the masks is minor. However, local biases can be significant as the maximum and minimum differences reach roughly $1\,\rm dex$, although the figure’s colour bar is clipped to $[-0.3, 0.3]$ for visualisation.

Figure~\ref{fig:d_z_compar} shows the redshift evolution of the ratios between the mean distance to the cosmic web of starbursts and the mean distance of a mass-matched sample of MS galaxies, and between quenched galaxies mass-matched to MS galaxies and the full MS galaxy sample when the reconstruction is performed without in-painting the masked areas. Compared to Fig.~\ref{fig:d_z_total}, the overall qualitative evolutions with redshift remain preserved, confirming that the large-scale behaviour is robust, but some differences are visible at $z<1$ for the starburst-MS ratio, and at $z>1$ for the quenched-MS ratio.

These figures provide a quantitative validation of the impact of the mask using an in-painting procedure described in Section~\ref{ssec:obs_cw}. Correcting for masked regions ensures that the measured distances to filaments are not artificially biased. However, the small sensitivity of our scientific results to the presence or absence correction demonstrates how robust our analysis is to these effects.

\section{Distance of galaxies to cosmic web nodes}\label{appD}

\begin{figure}[ht]
        \centering
        \includegraphics[width=\columnwidth]{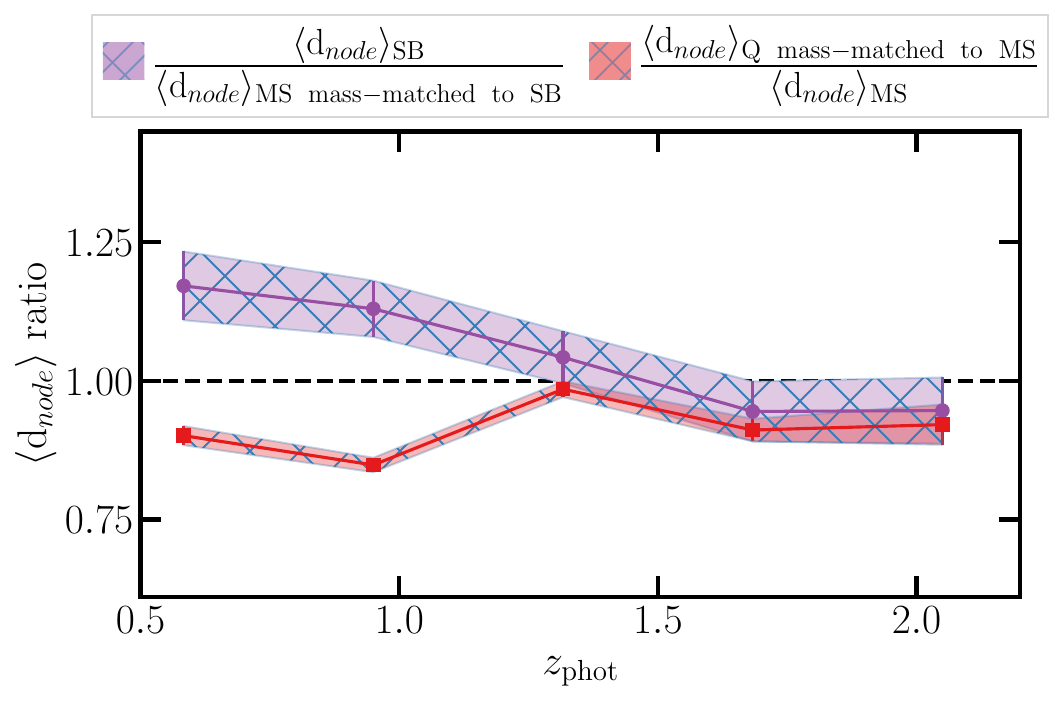}
        \caption{Same quantities as in the lower-left panel of Fig.~\ref{fig:d_z_total}, but for the distance to the nodes instead of the distance to the filaments.}
        \label{fig:d_z_nodes}
\end{figure}

For completeness, we repeated the analysis of Sect.~\ref{ssec:obs_res} with galaxies from the \cosmos~field using the \cosmosweb~data, but using the distance to nodes of the cosmic web $\mathrm{d}_{node}$ instead of the distance to filaments. The methodology is identical, and the results are displayed in Fig.~\ref{fig:d_z_nodes}, which shows the analogue of the lower-left panel of Fig.~\ref{fig:d_z_total}. 
In practice, we define nodes as critical points where filaments intersect, following the standard definition adopted by \disperse.

We find qualitative hints of a similar ordering of galaxy populations, with starbursts lying somewhat further away from nodes than mass-matched MS galaxies at low redshift, and quiescent systems showing the opposite trend. However, the relative offsets are modest, never exceeding the $\sim 15\%$ level, and are less striking than in the filament-based measurement. For filaments, the relative distance between starbursts and MS galaxies evolves from a ratio of $\sim 0.75$ at high redshift to $\sim 1.45$ at low redshift, corresponding to a variation of order $\Delta\dfil\sim 0.65$ over the redshift range considered. In contrast, when using the distance to nodes, the analogous ratio only varies from $\sim 0.9$ to $\sim 1.15$, i.e. a total variation of $\Delta\dfil\sim 0.2$. The amplitude of the evolution is therefore smaller by a factor of $\sim 3$ compared to the filament-based measurement. A similar but weaker behaviour is seen for quiescent galaxies relative to the main sequence, with ratios ranging from $\sim 0.95$ to $\sim 0.85$, although the trend is not monotonic with redshift. Overall, the node-based signal is significantly reduced in amplitude and less coherent than for filaments, which calls for cautious interpretation. 

In \simba, the node-based signal is also reduced when measured in projection, although it remains more pronounced than in \cosmos, and closer to the filament-based measurement shown in Fig.~\ref{fig:d_z_prediction_ratios}. This suggests that part of the attenuation is intrinsic to the 2D reconstruction of the cosmic web. In projection, nodes correspond to localised density peaks whose positions are more sensitive to redshift uncertainties and sampling noise than the more extended filamentary structures, leading to a stronger dilution of the environmental contrast. The additional weakening observed in \cosmos~may therefore reflect the combined effect of projection and more limited statistics.

Given the smaller amplitude of the effect, we refrain from drawing firm conclusions at this stage. These results are nevertheless consistent with the picture suggested by the filament analysis, and could be revisited in future work with larger samples or deeper data, with additional treatment of the effects of projection on nodes of the cosmic web.

\section{Control test with randomised FIR luminosities}\label{appE}

\begin{figure}[ht]
        \centering
        \includegraphics[width=\columnwidth]{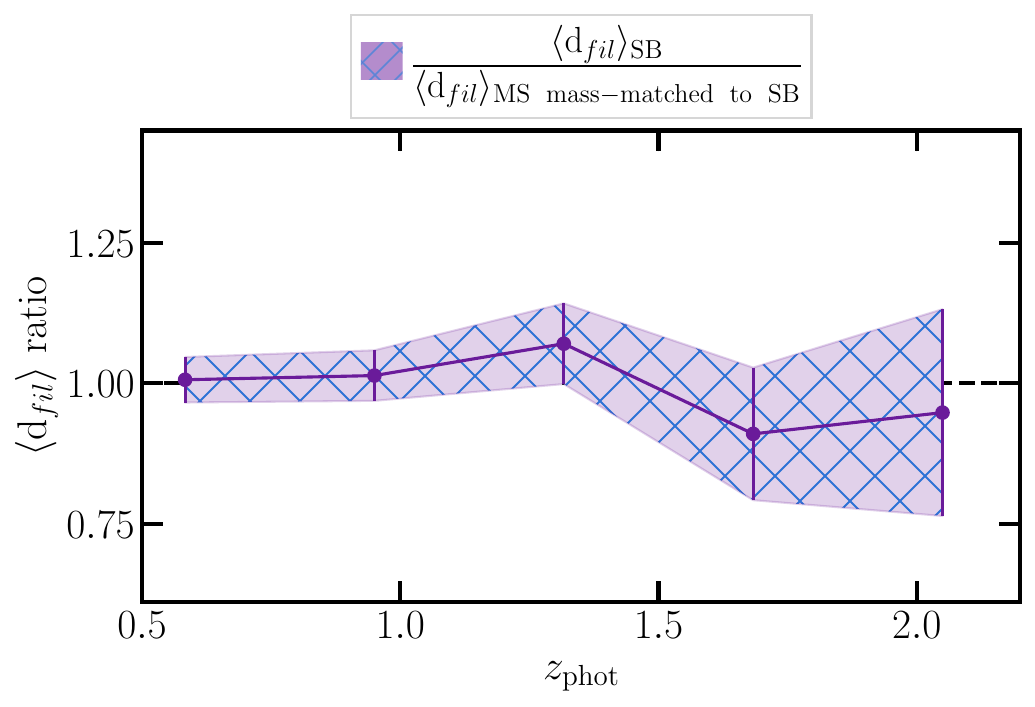}
        \caption{Same quantities as in the lower-left panel of Fig.~\ref{fig:d_z_total} for the starburst/MS ratio, but after averaging 1000 realisations of randomised $L_{\rm IR}$ values. This corresponds to selecting random galaxies from the fiducial catalogue, restricted to sources with ${\rm S/N} \geqslant 3$ in at least three bands, and repeating this process 1000 times, then taking the average ratio values.}
        \label{fig:d_z_random}
\end{figure}

To verify that the observed separation between starburst and MS galaxies is not driven by random fluctuations in $L_{\rm IR}$, we randomise the FIR luminosities within redshift bins and recompute the starburst/MS distance ratios. Figure~\ref{fig:d_z_random} shows the average over 1000 realisations. The flat trend confirms that the signal reported in Sect.~\ref{ssec:obs_res} arises from genuine correlations with the cosmic web rather than from random noise or fitting artefacts.

\section{Redshift-dependent modulation of sSFR by the distance to cosmic filaments}\label{appF}

\begin{figure}[ht]
        \centering
        \includegraphics[width=\columnwidth]{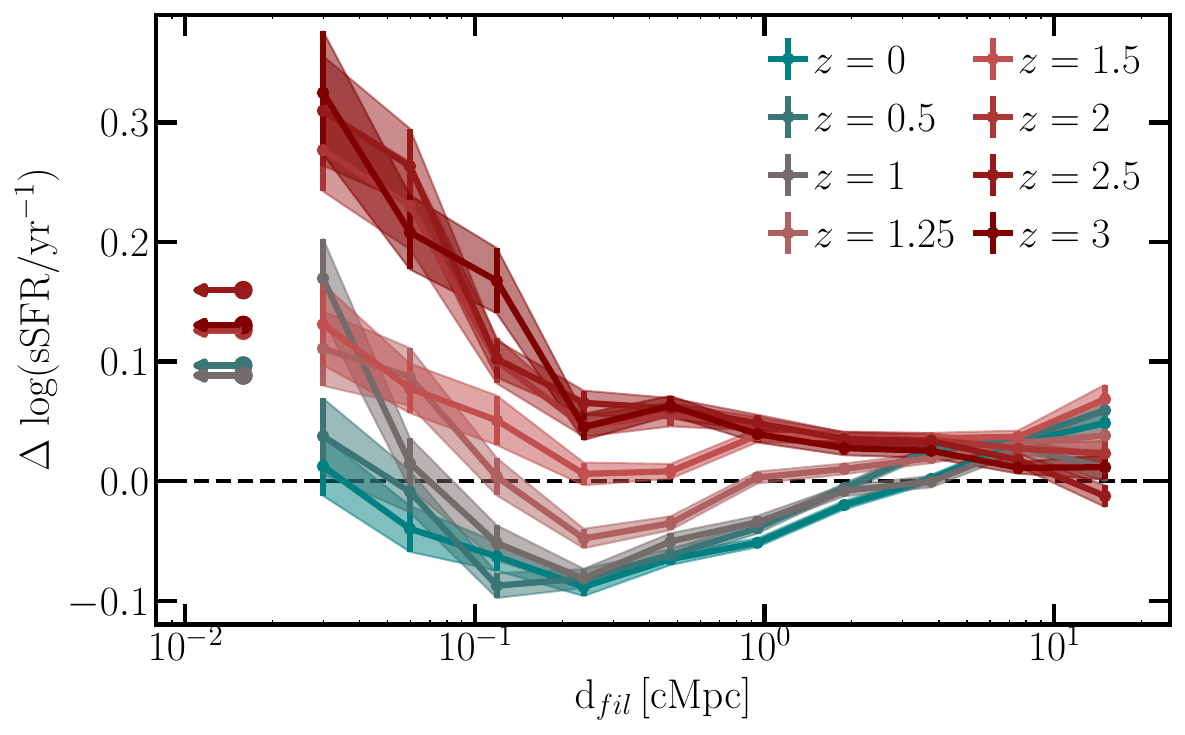}
        \caption{sSFR residuals as a function of the distance to the filaments after correcting for stellar-mass effects for SFG in \simba~in bins of distance. The arrows pointing to the left represent galaxies located at smaller distances than the plotted range, grouped into a single bin. Adapted from \citetalias{jego2026}.}
        \label{fig:DsSFR_dfil}
\end{figure}

We repeat the study presented in \citetalias{jego2026}, focusing on the redshift-dependent modulation of sSFR by the distance to cosmic filaments. In \simba, after selecting SFG as galaxies at each studied redshift $z$ with ${\rm sSFR} > 10^{-10 + 0.3\times \min(z, 2)}\,\rm yr^{-1}$, we fit a stellar-mass-dependent double power-law to the SFG sample only with
\begin{equation}
    {\rm sSFR}_{bf}(M_{\star}) = \frac{A}{(\frac{M_{\star}}{M_{\text{break}}})^{\alpha} + (\frac{M_{\text{break}}}{M_{\star}})^{\beta}}
    \label{eq:Qbf}
\end{equation}
with the amplitude $A$, the mass at slope break $M_{\text{break}}$, and the slopes $\alpha$ and $\beta$ as fitted parameters. At each redshift, we verify that this best-fit relation is representative of the SFG SFR-$M_{\star}$ distribution.
We then compute the mean deviation from this relation, $\Delta \log({\rm sSFR})$ in bins of distance to the filament as
\begin{equation}
    \Delta \log({\rm sSFR})_{\text{bin}} = \frac{\sum_{i=1}^{N_{\text{gal}}}\log\left(\frac{{\rm sSFR}_{i}}{{\rm sSFR}_{i, bf}} \right)}{N_{\text{gal}}},
    \label{eq:Dlog}
\end{equation}
where we sum over the number of galaxies in the bin $N_{\text{gal}}$. This provides a measure of the excess or deficit of sSFR beyond the effect of stellar mass. The values of $\Delta \log({\rm sSFR})$ as a function of the distance to filaments ${\rm d}_{fil}$ are given at various redshifts in Fig.~\ref{fig:DsSFR_dfil}. We use these trends to build the toy model presented in Sect.~\ref{ssec:intp_origin}, and to interpret the different evolutions of the average distance to filaments between starburst and MS galaxies observed in \cosmos~in Sect.~\ref{ssec:intp_intp}.

\end{appendix}

\end{document}